\documentclass[superscriptaddress,twocolumn,10pt,pra,aps,showpacs,longbibliography]{revtex4-2}

\usepackage{amsbsy}
\usepackage{amstext}
\usepackage{braket}
\usepackage[unicode=true,pdfusetitle,
 bookmarks=false,
 breaklinks=false,pdfborder={0 0 1},backref=false,colorlinks=false]
 {hyperref}
\hypersetup{
 bookmarksnumbered=false,bookmarksopen=false}

\usepackage{hyperref}
\hypersetup{colorlinks,citecolor=NavyBlue,filecolor=black,linkcolor=black,urlcolor=black}
\usepackage[dvipsnames]{xcolor}

\makeatletter

\@ifundefined{textcolor}{}
{%
 \definecolor{BLACK}{gray}{0}
 \definecolor{WHITE}{gray}{1}
 \definecolor{RED}{rgb}{1,0,0}
 \definecolor{GREEN}{rgb}{0,1,0}
 \definecolor{BLUE}{rgb}{0,0,1}
 \definecolor{CYAN}{cmyk}{1,0,0,0}
 \definecolor{MAGENTA}{cmyk}{0,1,0,0}
 \definecolor{YELLOW}{cmyk}{0,0,1,0}
}

\def\<#1>{\mathinner{\langle#1\rangle}}
\def\|#1>{\mathinner{|#1\rangle}}

\usepackage{amsmath}
\usepackage{graphicx}
\usepackage{amssymb}
\usepackage{txfonts,color}
\makeatother

\begin{document}

\title{Improving quantum state transfer: Correcting non-Markovian and distortion effects}
\author{Guillermo F. Pe{\~n}as}\email{guillermof.pens.fdez@iff.csic.es}
\affiliation{Instituto de F{\'i}sica Fundamental, IFF-CSIC, Calle Serrano 113b, 28006 Madrid, Spain}
\author{Ricardo Puebla}
\affiliation{Instituto de F{\'i}sica Fundamental, IFF-CSIC, Calle Serrano 113b, 28006 Madrid, Spain}
\affiliation{Departamento de F{\'i}sica, Universidad Carlos III de Madrid, Avda. de la Universidad 30, 28911 Legan{\'e}s, Spain}
\author{Juan Jos\'e Garc\'ia-Ripoll}
\affiliation{Instituto de F{\'i}sica Fundamental, IFF-CSIC, Calle Serrano 113b, 28006 Madrid, Spain}
\begin{abstract}
Quantum state transfer is a key operation for quantum information processing. The original pitch-and-catch protocols rely on flying qubits or single photons with engineered wavepacket shapes to achieve a deterministic, fast and high-fidelity transfer. Yet, these protocols overlook two important factors, namely, the distortion of the wavepacket during the propagation and non-Markovian effects during the emission and reabsorption processes due to time-dependent controls. Here we address both difficulties in a general quantum-optical model and propose a correction strategy to improve quantum state transfer protocols.  Including non-Markovian effects in our theoretical description, we show how to derive control pulses that imprint phases on the wavepacket that compensate the distortion caused by propagation. Our theoretical results are supported by detailed numerical simulations showing  that a suitable correction strategy can improve state transfer fidelities up to three orders of magnitude.
\end{abstract}
\maketitle

\section{Introduction}

Quantum state transfer, as originally conceived in Ref.~\cite{Cirac1996}, is the direct transport of quantum information between quantum processing units, through the exchange of flying qubits, in the form of  photons~\cite{Cirac1999, Xiang2017, Vogell2017}, magnons in spin chains~\cite{Yao2011}, or phononic and optomechanical excitations~\cite{Stannigel2010, Stannigel2011, Wang2011, Lemonde2018, Calajo2019, Vermersch2016}, to name some possibilities. Quantum state transfer can be used to distribute entanglement~\cite{Cirac1996} and, either directly or through protocols such as quantum teleportation~\cite{Bennet1993}, enable scalable quantum interconnects~\cite{Awschalom2021} for, e.g. quantum communication networks and distributed quantum computations (DQC)~\cite{Kimble2008,Wehner2018, Cacciapuoti2020} among multiple quantum processing units (QPUs).


The original pitch-and-catch protocols for quantum state transfer---see ~\cite{Cirac1996} and the above cited proposals---rely on flying qubits with engineered wavepacket shapes for deterministic, fast and high-fidelity emission and reabsorption in the source and destination QPUs. This shaping is achieved by tuning the interaction between the QPUs and the flying qubits, as demonstrated in experiments with superconducting circuits with microwave photons~\cite{Magnard2020, Kurpiers2017, Leung2019, Chang2020}, ions coupled to optical fibers~\cite{Ritter2012} and phonons~\cite{Bienfait2019}.

Unfortunately, the quantum-optical models for pitch-and-catch protocols neglect two important factors: (i) the distortion of photons in the quantum link and (ii) the non-Markovian effects induced by the time-dependent couplings. Regarding the first factor, Ref.~\cite{Penas2022} showed that the curvature of the dispersion relation in a quantum link sets stringent bounds on the ultimate state transfer fidelity. This affects potentially many experiments with microwave guides~\cite{Magnard2020, Kurpiers2017}, photonic crystals~\cite{Yu19,Butt21} or spin waves~\cite{ramos2016non, casulleras2022generation}. As for the non-Markovian effects~\cite{tufarelli2014non,laine2010measure}, these arise in the chirping of qubit-photon couplings, and manifest in distortions of the emitted wavepackets, which deviate from the predictions of the usual input-output theory. Note that both types of errors become more relevant as we increase the speed of state transfer and the bandwidth of the flying qubits. One solution is to exchange very narrow bandwidth photons, adopting the adiabatic limit of state transfer~\cite{Pellizzari1997,Chen2007,Ye2008,Clader2014, Tian2012}. However, this leads to slower protocols, subject to larger errors due to qubit decay and dephasing.



In this work we address both limitations simultaneously without compromising the speed of state transfer. We consider a fully-coherent transfer and solve the photon dispersion problem by imprinting the flying qubit with engineered phases and frequency chirps that compensate for the errors in the quantum link. On top of this, we design the emission and reabsorption of the flying qubits using an improved input-output model, with corrected equations that go beyond the Markov approximation. In this context we have quantified how much infidelity corresponds to the diffraction of the wavepacket and how much to the miscalibration of the emission process. We have verified numerically that our solution yields an improvement of over three orders of magnitude in the case of a microwave photon propagating in a superconducting waveguide with a non-linear dispersion relation.


This work is organized as follows. In Sec.~\ref{sec:model} we provide an overview of the considered system, i.e., a general and realistic quantum optical model describing the building block of a quantum network.  In Sec.~\ref{sec:control_design} we provide a theoretical analysis for the design of the control for wavepacket shaping. First, in Sec.~\ref{sec:correction_strategy} we describe the mechanism for wavepacket distortion, followed by Sec.~\ref{subsect:IO_relations} where we present the derivation of the control to yield a desired photon. In Sec.~\ref{sec:Beyond_Markov} we show how to take into account non-Markovian corrections. The theoretical analysis is then exemplified by detailed numerical simulations regarding an implementation in a superconducting architecture, presented in  Sec.~\ref{sec:results}. Finally, in Sec.~\ref{sec:summary} we summarize the most important results of the article and discuss possible future directions.



\begin{figure}
    \includegraphics[width=\columnwidth]{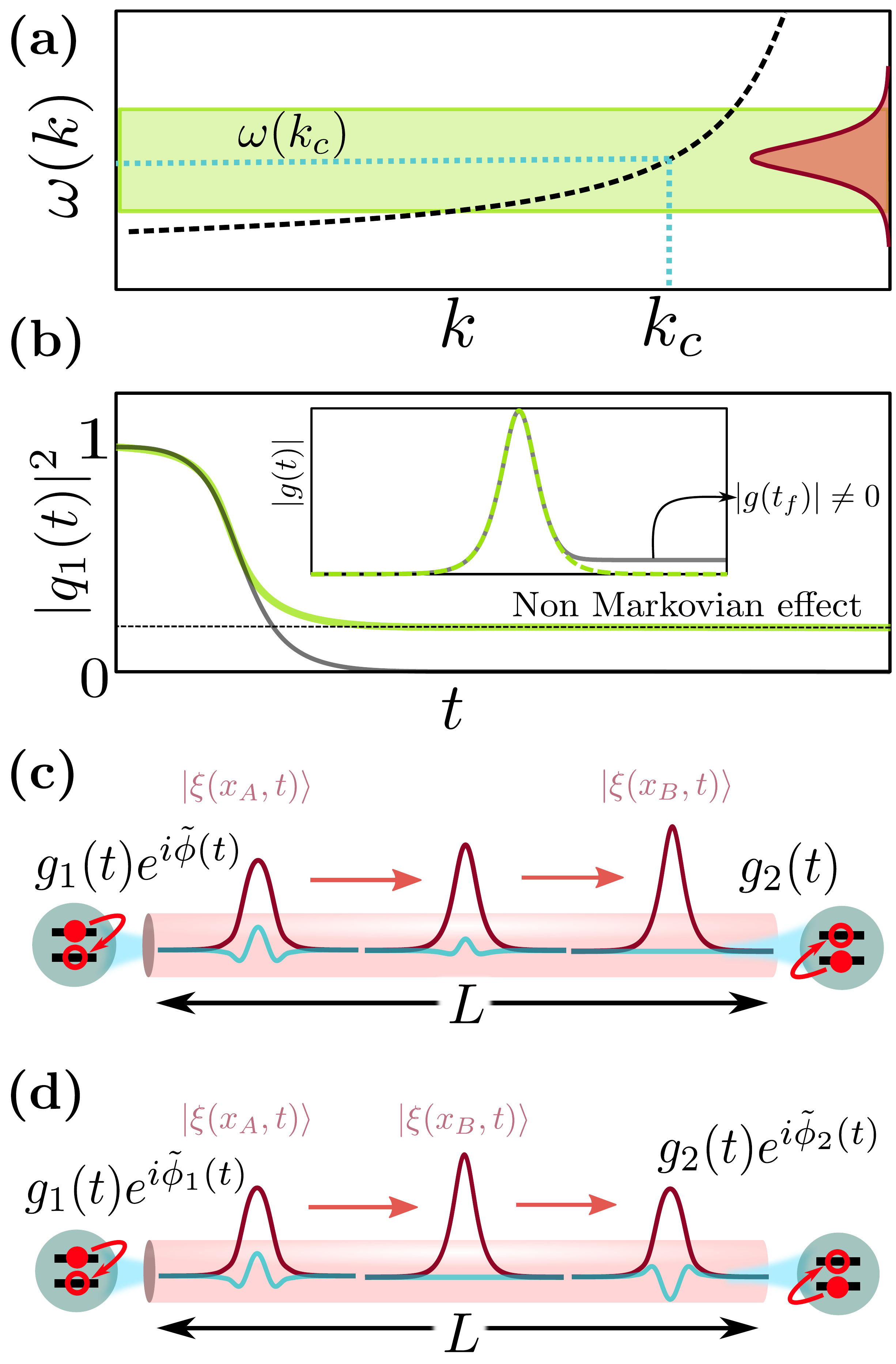}
    \caption{(a) Illustration of the two sources of imperfections that we tackle in this article, namely, that the fast generated photons (wide in frequencies) are affected by both the curvature of the dispersion relation of the material through which they propagate and the edges of the frequency band within they are contained.
    (b) Neglecting non-Markovian effects leads to an imperfect depletion of the qubit excitation (green line). This stems from the revivals as the photon hits  the end of the frequency band. In the inset  we show the controls $g(t)$ for the two different cases: when non-Markovian corrections are taken into account, the control remains non-zero at the end and the depletion is perfect (grey line).  Panels (c) and (d) show a sketch of the aim of this article, that is, to engineer complex controls, where both the phase and modulus are tuned dynamically. These control pulses yield photons robust to distortion that improve quantum state transfer. A phase is imprinted in these photons such that the propagation through the medium leads to a photon on the other end which can be absorbed with either a real control (cf. panel (c)) or with a complex control (cf. panel (d)).}
    \label{fig1}
\end{figure}

\section{Model}\label{sec:model}

We consider a minimal building block of a quantum state transfer architecture, consisting of two QPUs $j=1,2$, with a communication buffer of just one qubit, each connected to a common a quantum link with a different tuneable coupling $g_j(t)$~\cite{Pechal2013,Zeytinoglu2015}. In existing circuit-QED setups~\cite{Magnard2020, Kurpiers2017}, the quantum link is a microwave guide supporting bidirectional communication via photons of discrete frequencies $\omega(k)$, and the QPU-link connection is mediated by a filter resonator that helps preserve the qubit information. In general, the goal is to engineer the optimal controls $g_j(t)$ that enable the sending QPU, $j=1$, to create a well crafted photon that can be absorbed by at the receiving end, $j=2$. This goal can be achieved using other architectures, employing different bosonic links---e.g. photonic crystals or phononic waveguides---, with additional or strictly no transfer resonators, and yet with very minimal changes to the theory presented below.

The Hamiltonian of our minimal instance may be split into the following terms
\begin{align}\label{eq:H}
    H=H_{\rm QL}+\sum_{j=1,2} H_j+H_{j{\rm -QL}}.
\end{align}
The first three terms model the quantum link and both QPUs ($\hbar=1$)
\begin{align}
    H_{\rm QL}&=\sum_{k} \omega(k)b^\dagger_k b_k,\\
    H_{j}&=\delta_j \sigma^+_j\sigma_j^-+\Omega_{{\rm R}j} a^\dagger_j a_j+g_j(t)\left(\sigma^+_ja_j +{\rm H.c.}\right),
\end{align}
while the last one describes the interaction between the transfer resonators and the quantum link $H_{j{\rm -QL}}=\sum_k G_{k,j} \left(b_k^\dagger a_j+{\rm H.c.}\right)$. In this notation, $a_j^\dagger$ and $a_j$ are Fock operators for the transfer resonators, and $b_k^\dagger$ and $b_k$ are Fock operators for the waveguide modes with waveguide number $k$, which we constrain to a set of $N_{\rm modes}$ relevant modes. The Pauli matrices $\sigma_j^+=\ket{1}_j\!\bra{0}_j$ describe qubits with frequencies $\delta_j$ that are close to resonance with the transfer resonator modes $\Omega_{{\rm R}j}$. The index $k_c$ labels the waveguide mode whose frequency is closest to the photons we will generate  $\Omega_{{\rm R}j} = \delta_j \simeq \omega(k_c)$ (cf. Fig.~\ref{fig1}(a)). Note that this rotating-wave approximation model~\eqref{eq:H} is justified as long as $|\Omega_{{\rm R}j}-\delta_j|\ll |\Omega_{{\rm R}j}+\delta_j|$, and $|g_j|,|G_{k,j}|\ll \omega(k),\Omega_{{\rm R}j},\delta_j$.

A single traveling photon wavepacket may carry a qubit of information between QPUs, provided the shape of this wavepacket is engineered for perfect emission and absorption~\cite{Cirac1995}. The physics of this state transfer happens in a single-excitation subspace, which is exactly reproduced by the Wigner-Weisskopf wavefunction
\begin{equation}\label{eq:ansatz}
  \ket{\Psi(t)} = \left[\sum_{j=1,2}\left(q_j(t)\sigma_j^++c_j(t)a_j^\dagger\right)+\sum_k \psi_k(t)b_k^\dagger \right]\ket{\mathbf{0}}.
\end{equation}
This state describes a coherent superposition of an excitation in any of the qubits ($\sigma_j^+$), any of the transfer resonators ($a_j^\dagger$) or any of the waveguide modes ($b_k^\dagger$), on top of the vacuum state $\ket{\mathbf{0}}$. Eq.~\eqref{eq:ansatz} is a solution of the Schrödinger equation, provided the coefficients satisfy a set of linearly coupled ordinary differential equations
\begin{align}\label{eq:eom1}
    i\dot{q}_j(t)&=\delta_j q_j(t)+g_j(t) c_j(t),\\ \label{eq:eom2}
    i\dot{c}_j(t)&=\Omega_{{\rm R}j} c_j(t)+g^*_j(t) q_j(t)+\sum_k G_{k,j}\psi_k(t),\\ \label{eq:eom3}
    i\dot{\psi}_k(t)&=\omega(k) \psi_k(t)+\sum_{j=1,2}G_{k,j}c_j(t).
\end{align}

Although Eqs.~\eqref{eq:eom1}-\eqref{eq:eom3} are exact, they are too complicated for an analytical manipulation and optimal design of the control pulses $g_j(t)$. Instead, usual treatments of the state transfer problem work with a linearized, Markovian input-output theory~\cite{Gardiner1985} that can be exactly solved for simple wavepacket shapes~\cite{morin2019deterministic, gorshkov2007photon}. However, those models neglect two relevant physical phenomena, which are (i) the wavepacket distortion caused by curved dispersion relations $\omega(k)$, and (ii) the non-Markovian effects associated to broadband wavepackets and time-dependent qubit-photon couplings (cf. Fig.~\ref{fig1}(a) and (b)).  Indeed, as shown in Ref.~\cite{Penas2022} already the wavepacket distortion may limit the performance of state transfer in realistic waveguides, becoming the most dominant effect. In the following we describe an improved control theory that takes into account both limiting phenomena. This is done by a careful design of photons that are robust to distortion and an extension of the effective equations that includes corrections beyond the Markov approximation.

\section{Control design}\label{sec:control_design}

\subsection{Correction strategy}\label{sec:correction_strategy}

As a first step, let us address the distortion experienced by the photon that mediates the state transfer. As core strategy, our goal is to produce a wavepacket shape $\xi_\text{target}(x_B,t)$ at a target position $x_B$ in the quantum link's waveguide. We achieve this goal by engineering the wavepacket that is emitted at the source QPU, located at $x_A$, introducing a predistortion phase that compensates for the diffraction experienced when travelling from $x_A$ to $x_B$.

Our target wavepacket at $x_B$ may be decomposed as a linear combination of the waveguide modes
\begin{equation} \label{eq:packet_target}
  \xi_\text{target}(x_B, t) \sim \sum_\omega \psi_\omega^{(B)} e^{-i\omega t}.
\end{equation}
Following the literature, we assume the field profile follows a hyperbolic secant $\xi_\text{target}(x_B, t) \sim \text{sech}(\kappa t/2)$. If the dispersion relation was linear, $\omega(k)=v_g |k|$, we could achieve this profile by injecting a similarly shaped wavepacket $\xi(x_B, t) \propto \xi(x_A, t-t_\text{AB})$ at the entrance of the quantum link $x_A$, with $t_\text{AB}=|x_B-x_A|/v_g$. However, in presence of generic dispersion relations $\omega(k)$ we expect that shape of the wavepacket will change from $x_A$ to $x_B$ as depicted in Fig.~\ref{fig1}(c) and (d).

Without loss of generality, let us consider a semi-infinite line, with photons originating at $x_A=0$ and moving right-wards with momentum $k$. In this setup, the field at different positions along the line can be written as
\begin{equation} \label{eq:packet_everywhere}
    \xi(x, t) = \sum_k \psi_{\omega(k)} \exp \left\{i k x - i \omega(k) t\right\}.
\end{equation}
This equation relates the shape and phase of the injected wavepacket $\xi(x_A, t) =  \sum_k \psi_{\omega(k)}^{(A)} \exp\left\{- i \omega(k) t\right\}$ to the target wavepacket $\xi(x_B,t)$ at any other position and time. When inverting this relation we separate the phase accumulated by the dispersion relation $\omega(k)$ into linear and non-linear contributions
\begin{equation}
    \omega(k) = \omega(k_0) + v_g (k-k_0) +  \omega_\text{NL}(k)
\end{equation}
where $v_g = \partial \omega(k)/\partial k\big|_{k_0}$ is the group velocity, and $\omega_\text{NL}(k)$ the  aggregate of all the non-linear terms in $k$. After some manipulation, we arrive the first result in this manuscript. The fact that there exists a shape of the injected wavepacket~\eqref{eq:packet_target} which will produce the target photon shape at $x_B$
\begin{align}\label{eq:dist_packet}
\xi_\text{injection}(t) := \xi(x_A,t) = \sum_k \psi_{\omega(k)}^{(B)} e^{+i \omega_\text{NL}(k) t_\text{AB}}  e^{-i \omega(k) t}.
\end{align}
The phase correction in frequency space implies that the emitted photon no longer has the usual hyperbolic secant profile in position space at $x_A$. Instead, the original modulus and phase depends on the non-linear term $\omega_\text{NL}(k)$, and on the time $t_{AB}$ that it takes to travel between $x_A$ and $x_B$.

In the following subsection we will discuss how to implement the predistorted wavepacket in the qubit-waveguide setup. However, before discussing this physical implementation, we have to explain that there are two complementary strategies for how to use these correction schemes. The two strategies are depicted in Fig.~\ref{fig1}(c) and (d). In panel (c) the points $x_A$ and $x_B$ are chosen as the ends of the waveguide, and consequently, the photon at the reception point has an ideal hyperbolic secant profile. As we will see later on, this is not always feasible, specially if  either the separation $|x_B-x_A|$, or the non-linear terms $\omega_\text{NL}$ become so large, that the phase corrections cannot be implemented with the available coupling strengths. In this situation, we can apply the strategy from panel (d), where we engineer an ideal wavepacket at a point $B$ in the middle of the waveguide, and we apply correction strategies both at the injection and reception points.

\subsection{Input-Output relations} \label{subsect:IO_relations}

Let us discuss the time dependent control pulse $g(t)$ that implements our wavepacket correction strategies, studying an excitation that is transferred from a single qubit, to a semi-infinite waveguide, by means of the transfer resonator. Since the output field is the photon we wish to inject $\xi(x_A,t)$ is determined by our correction strategy~\eqref{eq:dist_packet}, the cavity field is also determined by Eq.~\eqref{eq:eom3} and we could in principle solve the remaining equations to determine the coupling $g(t)$ that produces this specific photon. In practice this is too difficult, because of the complexity of the equations, involving all waveguide modes, plus  the inhomogeneous time dependent term $g_j(t)q_j(t)$.

The standard solution to this problem is to perform the Markov approximation in the cavity's equation, replacing all cavity-waveguide coupling terms by the effective decay rate $\kappa$
\begin{align}\label{eq:effective_model_Markov}
    \dot{q}(t)&=-i g(t)c(t),\\ \label{eq:effective_model_Markov2}
    \dot{c}(t)&=-i g^*(t)q(t)-\kappa c(t)/2,
\end{align}
and tuning the qubit's frequency to the match the cavity, $\delta = \Omega + \delta\omega$ up to the Lamb shift $\delta\omega$. In addition to providing an equation for the qubit and cavity, the Markovian limit enables us to derive an input-output relation~\cite{GardinerUltracoldII}, which relates the shape of the photons injected into the waveguide at the position $x_A$, to the cavity field.
\begin{align} \label{eq:I-O relation}
c(t) = \xi(x_A,t)/\sqrt{\kappa}.
\end{align}

We can now invert these equations, recovering the complex control $g(t) \in \mathbb{C}$ from the wavepacket shape $\xi(x_A,t)$. Let us rewrite the cavity field as $d(t)=-ic(t)=e^{r(t)-i\theta(t)}$, and the complex qubit field as $q(t)=e^{x(t)-i \sigma(t)}$. Substituting these two relations into~\eqref{eq:effective_model_Markov}, we arrive at
\begin{align}\label{eq:control}
    g(t)=\frac{\dot{q}(t)}{d(t)}=\frac{(\dot{x}(t)-i\dot{\sigma}(t))e^{x(t)-i\sigma(t)}}{e^{r(t)-i\theta(t)}},
\end{align}
where every quantity on the right hand side is determined by the desired pulse shape---see App.~\ref{app:pulse_derivetion} for the details of the calculation. From Eq.~\eqref{eq:control} we see that the control $g(t)$ is in general complex. Fortunately, platforms of interest such as superconducting circuits~\cite{Zeytinoglu2015,Pechal2013} allow the engineering of rather general complex controls.

At this point, it is important to mention that not every wavepacket can be generated. Eq.~\eqref{eq:control} is only valid under the mathematical constraint $0\leq |q(t)|^2\leq 1 \ \forall t$, which implies an upper bound on the logarithmic derivative of the photon wavepacket (see App.~\ref{app:pulse_derivetion}). In physical terms, this sets a limit on the speed with which the wavepacket envelope may change, as well as a limit on the bandwidth of the photon and of the controls, both determined by $\kappa$. It also means that there exists a limit in the size of distortions $ \omega_\text{NL}(k) t_\text{AB}$ in~\eqref{eq:dist_packet} we can correct, given by the fastest rate at which $g(t)$ may be changed. As seen below, this maximum distortion can be quantified, providing insight on the design of actual experiments. Moreover, this limitation is the reason for the strategy in Fig.~\ref{fig1}(d), where the total correction is split between the emitting and the receiving points, to work around those limitations.

\subsection{Beyond Markov Model}
\label{sec:Beyond_Markov}

As we will see below, the strategy from Sec.~\ref{subsect:IO_relations} does not always produce a good fidelity of state transfer. Qualitatively, the correction of the distortion introduces rapid chirps in the coupling $g(t)$ and in the qubits' and cavities' dynamics, which cannot be fully captured by the Markov approximation introduced above---which assumes that the cavity field $c(t)$ is slowly modulated around a central carrier frequency $\sim\Omega$. Fortunately, we can trace back the quantum optical derivation of input-output theory, to introduce the lowest order corrections that consider a chirp in the modulation.

Let us begin with a standard quantum optical method~\cite{GardinerUltracoldII, ripoll2022quantum}, which is to solve formally for~\eqref{eq:eom3} and substitute the result into~\eqref{eq:eom2}. For our current purpose, which is to produce a well shaped photon, we can focus on a single qubit, resonator and semi-infinite waveguide. This results in an integro-differential equation
\begin{align}
       \dot{c}(t)=   -ig^*(t)q(t) +
        -i \xi_\text{in} (t-t_0)
       - \int_{t_0}^t d \tau  K(t-\tau) c(\tau),
\end{align}
where we have moved to a rotating frame at a frequency $\Omega_{R1}$, $\xi_\text{in}(t-t_0) = \sum_k G_k e^{-i \tilde{\omega}(k) (t-t_0)} \psi_k(t_0)$, $K(t-\tau) = \sum_k G^2_k e^{-i \tilde{\omega}_k( t - \tau)}$ is the kernel and $\tilde{\omega}(k) = \Omega_{R1}-\omega(k) $ is the detuning between the cavity and each of the waveguide modes.

A standard state transfer experiment starts with no field in the waveguide, $\psi_k(t_0) = 0$. Thus, we thus can write
\begin{align}\label{eq:cavity}
    &\dot{c}(t) =  -ig^*(t)q(t) - \Gamma(t).
\end{align}
The source term $\Gamma(t) = \int_{t_0}^t d \tau  K(t-\tau) c(\tau)$ may be expanded in increasing order of the derivatives of $c(t)$ (see App.~\ref{app:nonMArkov})
\begin{align}\label{eq:source_term}
  \Gamma(t) \approx c(t)  \int_{t_0}^t \!\mathrm{d}\tau  K(t-\tau) + \dot{c}(t) \int_{t_0}^t \!\int_{t_0}^\tau \!\!\mathrm{d}\tau \mathrm{d}\tau' K(t-\tau') + \cdots
\end{align}
In this work we only need the first two terms of the expansion
\begin{align}\label{eq:source_term2}
     \Gamma(t) \approx c(t) \left(i\delta\omega(t) + \frac{\kappa(t)}{2}\right) + \dot{c}(t)\mathcal{N}(t),
\end{align}
which are functions of the Lamb shift and decay rate $i\delta\omega(t) + \kappa(t)/2 = \int_{t_0}^t d \tau  K(t-\tau)$, and the correction $\mathcal{N}(t) = \int_{t_0}^t d\tau \int_{t_0}^\tau d\tau' K(t-\tau')$. Therefore, non-Markovian effects appear as a time-dependent decay rate $\kappa(t)$ and Lamb shift, as well as a first-order correction $\mathcal{N}(t)\neq 0$.

We will show below, that a good approximation to design the controls is to take average values $\kappa$ and $\mathcal{N}$ over the duration of the experiment. Using these constant values, we can solve the equation of motion for the cavity field
\begin{align}\label{eq:effective_model_non_Marvov}
    \dot{c}(t)&=\frac{1}{1-\mathcal{N}}\left(-i g^*(t)q(t)-\kappa c(t)/2\right),
\end{align}
where we assume once more the resonance condition $\delta=\Omega+\delta\omega$. Using this equation and the ideas in Sec.~\ref{subsect:IO_relations} we can now derive the optimal control $g(t)$ up to the first correction beyond the Markov approximation (see App.~\ref{app:pulse_derivetion}).

\section{Results}\label{sec:results}

Our theoretical developments have produced two central results. First, we have found that non-Markovian effects are relevant and that no efficient diffraction correcting protocol can be implemented without taking such non-Markovian effects into consideration. Second, the modified non-Markovian model allows for the faithful engineering of wavepackets to correct distortion, thanks to the ability of imprinting on-demand phases in the photons we aim to create. We will now confirm these results using numerically exact simulations of a specific and realistic implementation of the state transfer setup with circuit-QED components.

\subsection{Simulation Parameters}

In this implementation we assume a rectangular superconducting WR90 waveguide, which is commercially available and of widespread use in circuit-QED experiments. Such waveguide has a dispersion relation of the form
\begin{align}\label{eq:disp_rel}
    \omega(k_m)=c\sqrt{\left(\frac{\pi}{l_1}\right)^2+k_m^2},
\end{align}
where the $k_m$'s are the wave numbers in the long dimension defined as $k_m=m\pi/L$ ($x\in[0,L]$)~\cite{Pozar} and $l_1$ is the width of waveguide. From the theory of capacitive interactions between the transfer resonators and the cavities, we can deduce the cavity-photon coupling terms $G_{m,j}=(-1)^{m(j-1)}\sqrt{\kappa_j v_g \; \omega(k_m)/(2\Omega_{\text{R}j} L)}$ as a function of the experimental decay rates $\kappa_j$ of the two transfer resonators.

The numerical design and modelization of the experiments involves two components. First, we had to create a software to obtain the time-dependent controls $g_j(t)$. In some simple limits, such as  a waveguide with linear dispersion relation, this code can rely on analytical formulas for $g(t)$ (see App.~\ref{app:pulse_derivetion}). However, we solve the general problem using a Runge-Kutta method to integrate~\eqref{eq:effective_model_non_Marvov} backwards in time, starting from a desired predistorted wavepacket.

The second component in this work is a full simulation of the complete dynamical Eqs.~\eqref{eq:eom1},~\eqref{eq:eom2} and~\eqref{eq:eom3}. This solver uses a Trotterization of the dynamics that works for general dispersion relation, generic time-dependent controls and any free spectral range (FSR). These simulations provide numerically exact results within the single-excitation subspace, with which we benchmark the performance of the state transfer protocols.

In the following we discuss two different sets of simulations. Following \cite{Magnard2020} we choose a $L=5$m waveguide containing $351$ modes; and also a $L=60$m  waveguide containing $4000$ modes. The disparity in problem sizes is due to the scaling of the FSR with the inverse of the waveguide length, which requires more modes for a $60$m waveguide to account for similar bandwidth as in the $5$m experiment. In both cases we push the experiments to the the fastest available speeds, utilizing almost the whole X-band of the WR90 waveguide~\cite{Pozar}.

\subsection{Non-Markovianity and faithfulness of the effective model}

In this section we study the dynamics of the source term $\Gamma(t)$, as obtained from the numerically exact dynamics of $c(t)$. 

We focus on only one node, start with the qubit initially populated $q(0) = 1$ and inject a broadband hyperbolic secant photon with $\kappa/(2\pi) = 200$ MHz into a $L=5$m waveguide. This is an experimentally demanding limit, which allows for very fast protocols and a deep exploration of the non-Markovian corrections.

\begin{figure}[h!]
    \includegraphics[width=\columnwidth]{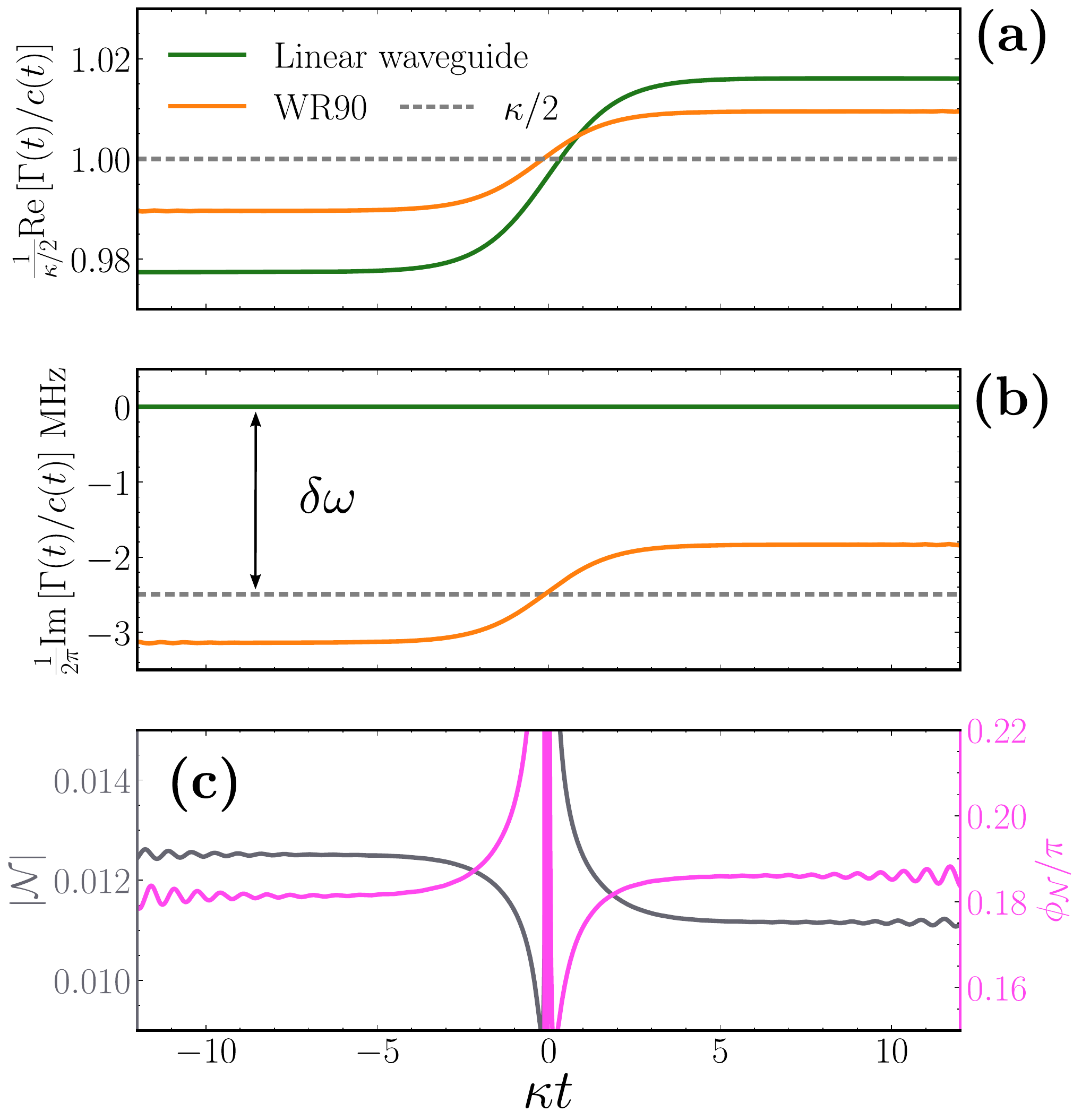}
    \caption{Real (a) and imaginary (b) part of $2/\kappa \ \Gamma(t)/c(t)$ for a linear waveguide with constant couplings $G_k = |G|$ (green line) and for a WR90 model (orange line). The real part gives account of the decay rate, while the imaginary part refers to a Lamb shift. (c) Modulus and phase of the non-Markovian parameter $\mathcal{N} = |\mathcal{N}| e^{i \phi_\mathcal{N}} $ calculated from the two top panels according to Eq.~\eqref{eq:non_Mark_param}. Parameters: $L=5$m  waveguide, $\kappa/(2\pi) = 200$ MHz, 5000 time steps, $g(t) \propto \rm sech(\kappa t) $.}
    \label{fig2}
\end{figure}

Using these parameters, we solve the complete dynamics of the link and analyze the ratio $\Gamma(t)/c(t)$, shown on the top two panels of Fig.~\ref{fig2}. Note that in a purely Markovian theory, this ratio would be given by a constant complex number $i\delta\omega +\kappa/2$, whose real and imaginary parts give the cavity's decay rate and effective Lamb shift. For a linear dispersion relation we find that the Lamb shift does indeed vanish (there is no imaginary component, as in the Markovian theory), but the ratio $\Gamma(t)/c(t)$ is not constant. Instead, the ratio $\mathrm{Re}(\Gamma(t)/c(t))$ moves around the designed decay rate $\kappa$, staying below and above, when $\dot{c}(t)/c(t) >0$ and $\dot{c}(t)/c(t)<0$, respectively. Similarly, when considering a curved dispersion relation, we find that the average Lamb shift is no longer zero, but it also experiences similar deviations that can be explained by a term proportional to $\dot{c}(t)/c(t)$.

\begin{figure}[h!]
    \includegraphics[width=\columnwidth]{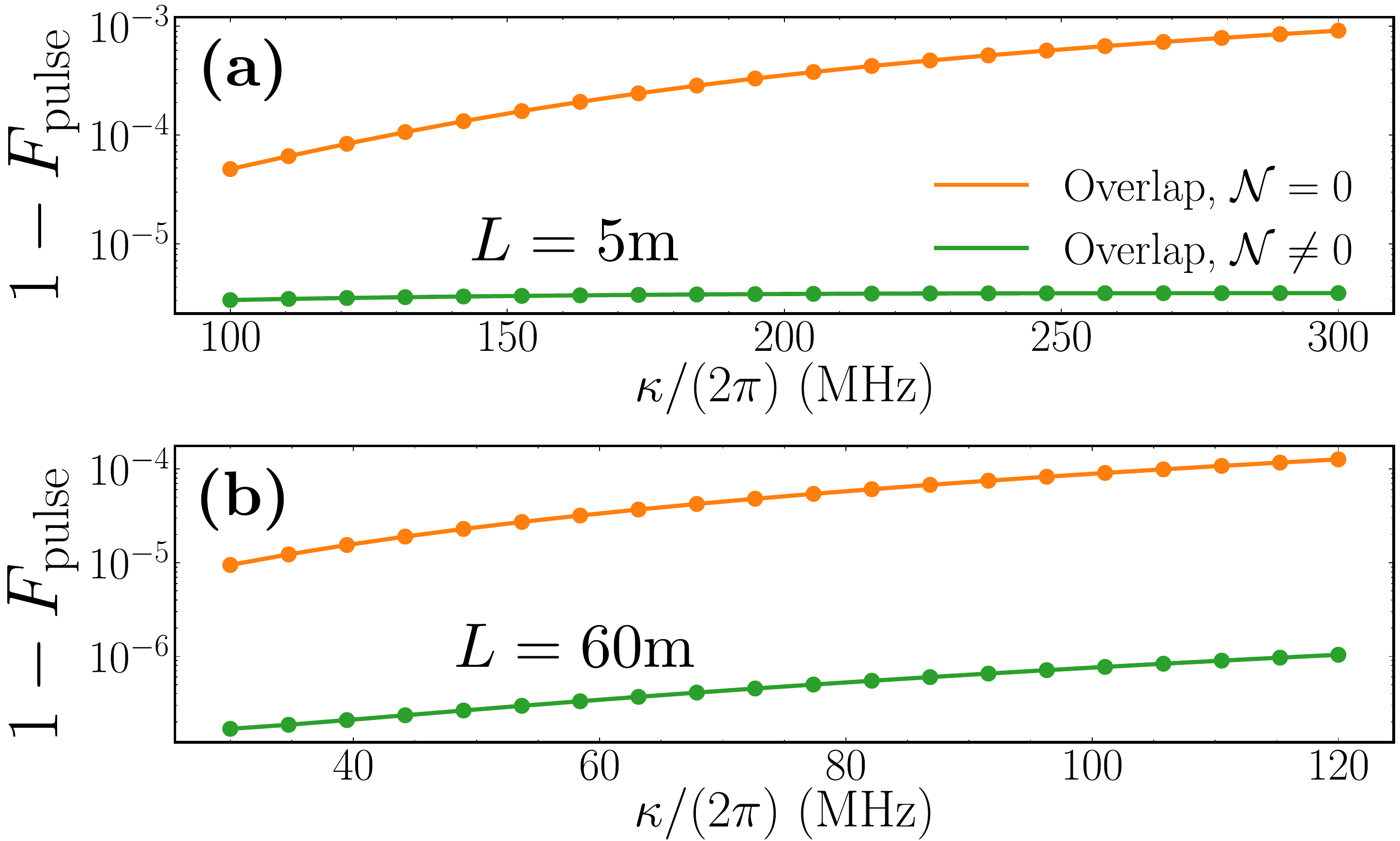}
    \caption{Infidelity $1-F_{\rm pulse}$ to quantify the faithtulness of the effective model for two different waveguide lengths $L=5$ m (a) and $L=60$ m (b). The control $g(t)$ is derived from a Markov model ($\mathcal{N}=0$, orange lines) and from the effective model including non-Markovian effects ($\mathcal{N}\neq 0$, green lines). The target field in every case is $\psi_\text{target}(t)=\sqrt{\kappa/4} \; {\rm sech}(\kappa t/2)$.  The output field has been reconstructed at the middle of the waveguide by means of~\eqref{eq:packet_everywhere} setting $x=L/2$. For $L=5$ m, 351 modes, 2000 time steps have been used and a protocol duration of $t_\text{5m} \in (-40/\kappa, +40/\kappa)$. For $L=60$ m,  4000 modes, 2000 timesteps and $t_\text{60m} \in (-100/\kappa, +100/\kappa)$.}
    \label{fig3}
\end{figure}

Using the computed values of $c(t)$ and $\Gamma(t)$ we may estimate the strength of the non-Markovian correction $\mathcal{N}(t)$. For that we assign $\kappa$ and $\delta\omega$ their averaged values (grey dashed lines in Fig.~\ref{fig2}(a) and (b)), and solve
\begin{align}\label{eq:non_Mark_param}
    \mathcal{N}(t) =  \left(  \frac{\Gamma(t)}{c(t)} - \kappa/2 -i \delta \omega \right) \frac{c(t)}{\dot{c}(t)}.
\end{align}
Fig.~\ref{fig2}(c) shows the estimated correction for an emission process  running from $t_\text{emission} \in (- 12/\kappa, + 12/\kappa)$. Around $t=0$, the population in the cavity reaches a maximum and therefore the term $c(t)/\dot{c}(t)$ in equation~\eqref{eq:non_Mark_param} becomes singular, invalidating the above expression. However, for most of Fig.~\ref{fig2}(c), we find $\mathcal{N}(t)\approx \mathcal{N}$ is approximately a constant complex number, with fixed amplitude and phase.

The deviations from the Markovian theory have practical consequences. They mean that a control $g(t)$ designed in the purely Markovian limit~\eqref{eq:effective_model_Markov2} will produce photons $\xi_\text{injection}$ that fail to meet the desired profile $\xi_\text{target}$. These errors are reduced if we instead use the non-Markovian equations with a nonzero $\mathcal{N}$. We have tested this in practice, computing the fidelity of injected photons
\begin{align}
  F_\text{pulse} = |\braket{\xi_\text{target}| \xi_\text{injection}}|^2,
\end{align}
as obtained from simulations of the complete Hamiltonian, with controls $g(t)$ derived both in the Markovian and corrected theories. The results are shown in Fig.~\ref{fig3} for a $L=5$m and a $L=60$m waveguides with linear dispersion relation. Already in this simple scenario, in which the control $g(t)$ can be found analytically even for $\mathcal{N} \neq 0$ (see App.~\ref{app:pulse_derivetion}), we find that including the non-Markovian term $\mathcal{N}$ results in an improvement of over one order of magnitude in the fidelity of the pulses. This supports our claim that when including the beyond Markov correction, the resulting model describes the physics of the full system more faithfully, and empowers us to design a better state transfer protocol in a realistic scenario.

\subsection{Correcting distortion in state transfer}

In this subsection we will show how our method improves the quantum state transfer fidelity by several orders of magnitude employing a realistic WR90 waveguide. The chosen lengths are inspired by two recent experiments~\cite{Kurpiers2017}.
As explained in Sec.~\ref{sec:control_design}, in order to fix the distortion one needs to generate a wavepacket as given in Eq.~\eqref{eq:dist_packet}. We will expand the dispersion relation~\eqref{eq:disp_rel} up to second order in $(k-k_c)$
\begin{align}
    \omega(k) = \omega(k_c) + (k-k_c)v_g + \frac{1}{2}(k-k_c)^2 D_2,
\end{align}
where $D_2 =   \frac{\partial^2 \omega(k)}{\partial k^2}\big|_{k_0}$ is the curvature of the dispersion relation.

We consider as our ideal wavepacket one that is popular in the literature $\psi^\text{id}_k = \sqrt{\frac{\pi}{2\kappa}}{\rm sech}\left(\frac{\pi\omega(k)}{\kappa} \right)$. In order to correct the dispersion up to $D_2$, we must inject a photon wavepacket with the following phase profile
\begin{align} \label{eq:corrected photon}
    \xi_\text{injection}(t) = \sum_k \psi_k^\text{id}  e^{-i \omega(k) t}  \exp{\left(i \frac{1}{2} D_2 (k-k_0)^2 t_\text{AB}\right)}.
\end{align}
In the following we introduce a dimensionless distortion strength, given by $D = D_2 t_\text{AB}/(2v_g^2)$. From our numerical experiments, we find that the phase profile~\eqref{eq:corrected photon} may be produced by a suitable control $g(t)$, provided $|D|\leq D_{\rm max}$  where $D_{\rm max}\approx 3/(2\sqrt{5}\kappa^2)$ (see App.~\ref{app:max_distortion}).

\begin{figure}[t]
    \includegraphics[width=\columnwidth]{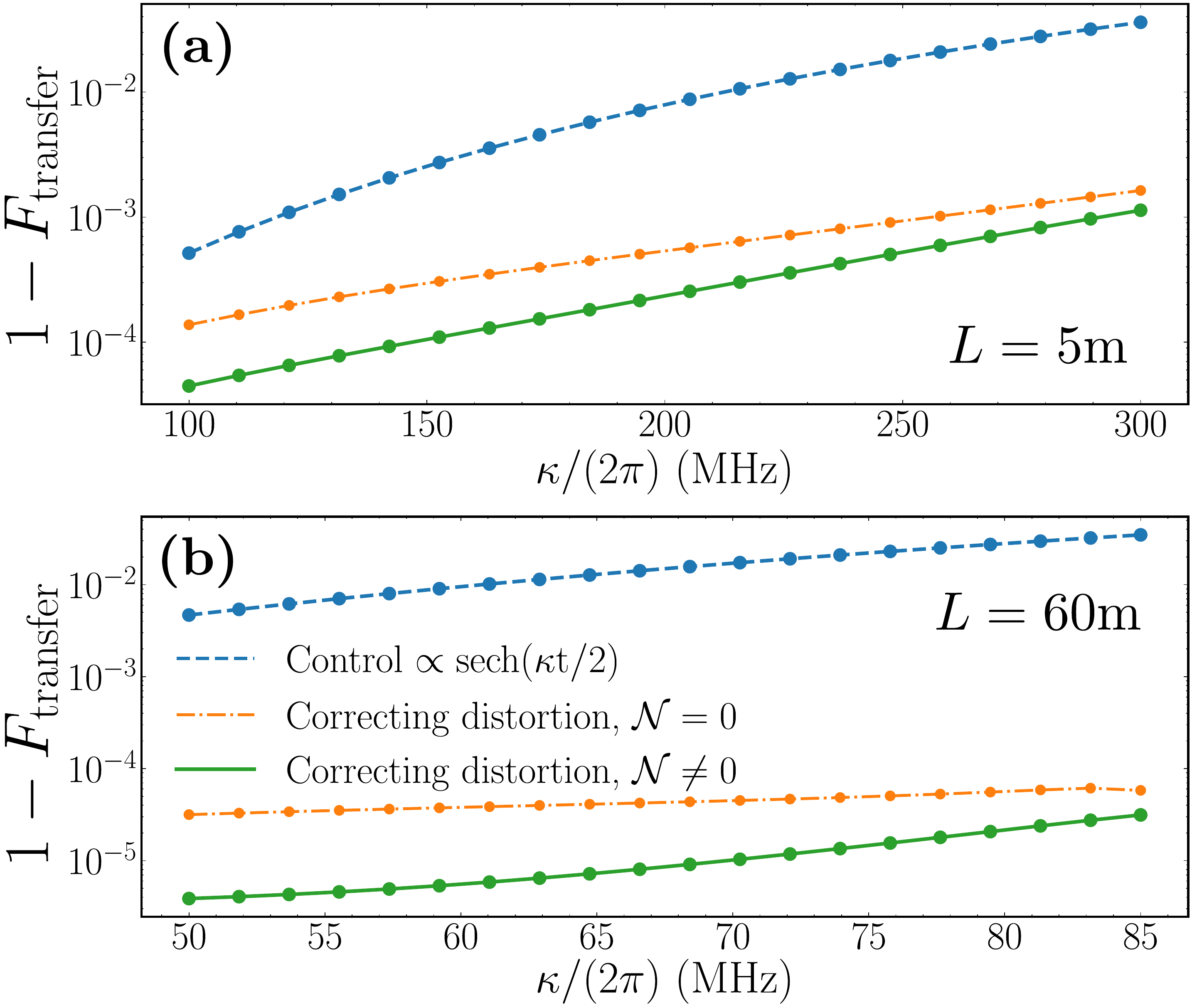}
    \caption{ State transfer infidelity for a realistic WR90 waveguide. The dots are the actual points where the simulation was performed, where as the lines are to be interpreted merely as guides to the eye. Blue lines correspond to the usual $g(t) =\kappa /2\; {\rm sech}(\kappa t/2)$. Orange lines are obtained by trying to generate the corrected photon~\eqref{eq:corrected photon}, obtaining the corresponding $g(t)$ with the effective Markovian model ($\mathcal{N}=0$). Green lines correspond to the infidelity when the control to generate~\eqref{eq:corrected photon} is obtained considering corrections beyond Markov ($\mathcal{N} \neq 0$). Both the blue and green lines are explained very accurately by an analytical calculation of the distortion, analogous to the one presented in \cite{Penas2022}. Panel (a) contains simulations for a 5m waveguide whose parameters are 351 modes, 2000 time steps and a protocol duration of $t_\text{5m} \in (-40/\kappa, +40/\kappa)$. A distortion of $D_\text{5m} = 0.33/2 \; \text{ns}^{-2}$ at each of the nodes. The parameters of panel (b) are 60m waveguide 4000 modes, 2000 time steps and $t_\text{60m} \in (-100/\kappa, +100/\kappa)$.  A distortion of $D_\text{60m} = 4.70/2 \; \text{ns}^{-2}$ at each of the nodes.}
    \label{fig4}
\end{figure}

Considering the parameter regime shown in Fig.~\ref{fig4},  i.e. $L=5$ m and $\kappa/(2\pi)\sim 10^2$ MHz and a $L=60$ m waveguide with $\kappa/(2\pi)\sim 10^1$ MHz, the values for the distortion parameter are $D_\text{5m} = 0.33 \; \text{ns}^{-2}$ and $D_\text{60m} = 4.70 \; \text{ns}^{-2}$. These values are larger than $D_{\rm max}$ but still smaller than $2D_{\rm max}$. Hence, one can employ the complex control to correct half the distortion on the emission process and half on the absorption, as discussed at the end of Sec.~\ref{sec:correction_strategy}.

In Fig.~\ref{fig4} we show how our method compares to the usual wavepacket engineering done ignoring non-linear dispersion relations. We define the fidelity of state transfer as the probability that the second qubit absorbs the photon and gets excited at the end of the experiment $t_f$
\begin{align}
    F_\text{transfer} = |q_2(t_f)|^2.
\end{align}
The blue line corresponds to the state transfer fidelity using the usual Markovian control $g(t) =\kappa /2\; {\rm sech}(\kappa t/2)$~\cite{Penas2022}, obtained from an ideal wavepacket such as $\psi^\text{id}_k = \sqrt{\frac{\pi}{2\kappa}}{\rm sech}\left(\frac{\pi\omega(k)}{\kappa} \right)$ without any predistortion of the phase. As expected, we obtain fidelities that decrease with increasing photon bandwidth. These infidelities are dominated by the photon wavepacket distortion, consistent with the estimations in~\cite{Penas2022}.

The orange and green lines show the fidelity obtained using controls that have been obtained proposing the $\xi_\text{injection}$ defined in Eq.~\eqref{eq:corrected photon}. Both of them show much better fidelities than the case of the ideal photon. The orange line uses time-dependent controls obtained with the Markovian effective model, cf. Eqs.~\eqref{eq:effective_model_Markov} and~\eqref{eq:effective_model_Markov2}. The values to which these line saturates can be explained by the orange line in Fig.~\ref{fig3}. Beware that the results of Fig.~\ref{fig3} can not be directly translated to Fig.~\ref{fig4}, since $G_k = |G|$ in the first one, whereas $G_k \propto \sqrt{\omega_k}$ in the second. Finally, the generation of Eq.~\eqref{eq:corrected photon} with a control $g(t)$ taking into account non-Markovian effects, with $\mathcal{N} \neq 0$, leads to infidelities that are up to three orders of magnitude better than with the usually employed controls. 
As for the blue line, we stress again that the numerical results for the green line are accurately predicted by the theoretical model proposed in~\cite{Penas2022} provided that one accounts for the correction of the quadratic term.

\section{Conclusions and outlook} \label{sec:summary}

In this work we have developed a new theoretical approach to optimizing quantum state transfer in realistic scenarios. The main result of the article is the demonstration that it is possible to achieve nearly optimal quantum state transfer even when photons suffer coherent distortions in the quantum link. To compensate for the distortions that appear in a medium with non-linear dispersion relations, we derived a new set of control equations that both address the need of complex couplings, while introducing corrections beyond the usual Markovian theory. Both ingredients combine to reduce the infidelity of quantum state transfer for several orders of magnitude in scenarios of practical experimental interest, such as circuit-QED experiments

This work opens up different possible avenues for continuation. On a very basic front, it may be interesting to understand how our theory applies to more general platforms, from photonic crystals to magnonic waveguides. More generally, we envision straightforward continuations that include higher order corrections to the Markovian limit, of which we have evidence already in these simulations. However, the most ambitious goal would be to push the speed limits of quantum state transfer, seeking interaction strengths and control speeds that violate the constraints of our model, such as the rotating-wave approximation.

\begin{acknowledgements}
We thank Alvaro G\'omez-Le\'on for valuable feedback on the manuscript. This work has been supported by the European Union's Horizon 2020 FET-Open project SuperQuLAN (899354) and Proyecto Sin\'ergico CAM 2020 Y2020/TCS-6545 (NanoQuCo-CM).
\end{acknowledgements}

\appendix

\section{Derivation of the control in the non-Markovian regime}\label{app:pulse_derivetion}
Here we provide the details of the derivation of $g(t)$ from the dynamical equations for the qubit and cavity field, namely,
\begin{align}
    \dot{q}(t)&=-i g(t)c(t)\\
    \dot{c}(t)&=\frac{1}{1-\mathcal{N}}\left(-i g^*(t)q(t)-\kappa c(t)/2\right),
\end{align}
we will assume now that $\mathcal{N} \in \mathbb{C}$, moreover, $\mathcal{N} = |\mathcal{N}| e^{i \phi_\mathcal{N}}$. Because all the quantities at play are complex, we have to keep track of both modulus and phase: $d(t)=-ic(t)=e^{r(t)-i\theta(t)}$, $q(t)=e^{x(t)-i \sigma(t)}$. We also make use of the input-output relation presented in the main text $d(t) =\xi(t)/\sqrt{\kappa}$, where $\xi(t) := \xi(x_A,t)$ for simplicity. That way we have
\begin{align}
 \frac{d}{dt}|q(t)|^2+(1-\mathcal{N})d^*(t)\frac{d}{dt}d(t)+(1-\mathcal{N}^*)d(t)\frac{d}{dt}d^*(t)=-\kappa |d(t)|^2.
\end{align}
Upon integrating the previous equation, we find
\begin{align} \label{eq:qubit_field}
 &|q(t)|^2 = |q(t_0)|^2 - \frac{(1-\mathcal{N})}{\kappa}\int_{t_0}^t \xi^*(\tau)\dot{\xi}(\tau) d\tau + \notag\\& - \frac{(1-\mathcal{N}^*)}{\kappa} \int_{t_0}^t \xi(\tau)\dot{\xi}^*(\tau) d\tau - \int_{t_0}^t |\xi(\tau)|^2 d\tau.
\end{align}
Now we focus on the equations for the phases. From
\begin{align}
    \dot{q}(t)=[\dot{x}(t)-i\dot{\sigma}(t)]q(t)&=g(t)d(t)\\
    \dot{d}(t)(1-\mathcal{N})=[\dot{r}(t)-i\dot{\theta}(t)]d(t)(1-\mathcal{N})&=-g^*(t)q(t)-\kappa d(t)/2,
\end{align}
we find
\begin{align}\label{eq:btaeq}
    d(t)=\frac{-g^*(t)q(t)}{(\dot{r}(t)-i\dot{\theta}(t))(1-\mathcal{N})+\kappa/2}.
\end{align}
On the other hand,
\begin{align}
    \dot{x}(t)-i\dot{\sigma}(t)=\frac{g(t)d(t)}{q(t)}.
\end{align}
Inserting Eq.~\eqref{eq:btaeq} into the previous expression we get
\begin{align}
    \dot{x}(t)-i\dot{\sigma}(t)=\frac{-|g(t)|^2}{(\dot{r}(t)-i\dot{\theta}(t))(1-\mathcal{N})+\kappa/2}.
\end{align}
Isolating real and imaginary parts, and recalling $\mathcal{N}=|\mathcal{N}|e^{i\phi_\mathcal{N}}$, we arrive to
\begin{align}
    \dot{\sigma}(t)=-\dot{x}(t)\frac{\dot{\theta}(t)(1-|\mathcal{N}|\cos\phi_\mathcal{N})+\dot{r}(t)|\mathcal{N}|\sin\phi_\mathcal{N}}{\dot{r}(t)(1-|\mathcal{N}|\cos\phi_\mathcal{N})-\dot{\theta}(t)|\mathcal{N}|\sin\phi_\mathcal{N}+\kappa/2},
\end{align}
and with $\sigma$ one has all the ingredients to plug in equation~\eqref{eq:control} of the main text and construct the desired control.

\subsubsection{Simple example}
A simple limiting case would be to emit a photon that has no imaginary part (for example, if we planned to correct all distortion at reception). Considering such a real pulse $\theta(t)=\sigma(t)=0 \ \forall t$, then
\begin{align}
    g(t)=\frac{\dot{x}(t)e^{x(t)}}{e^{r(t)}},
\end{align}
where $r(t)=\log (|\xi(t)|/\sqrt{\kappa})$ and $x(t)=\log|q(t)|$. For $\xi(t)=\sqrt{\kappa/4}{\rm sech}(\kappa t/2)$, we find
\begin{align} \label{eq:modified_control}
    g(t)=\frac{e^{\kappa t/(2-2\mathcal{N})}\kappa }{(1+e^{\kappa t/(1-\mathcal{N})})^2\sqrt{\mathcal{N}+(1-\mathcal{N})/(1+e^{\kappa t/(1-\mathcal{N})})^2 }}.
\end{align}
For $\mathcal{N}=0$ we recover the standard pulse $g(t)=\kappa/2 {\rm sech}(\kappa t/2)$. However, for $\mathcal{N}>0$ we find that $|q(t\rightarrow \infty)|^2=\mathcal{N}$.

\section{Non-Markovian regime}\label{app:nonMArkov}

Using the memory kernel, one can write the term $\Gamma(t)$ as
\begin{align}
    \Gamma(t)=\int_{t_0}^t d\tau \ \sum G_k^2 e^{-i\tilde{\omega}(k)(t-\tau)}c(\tau)=\int_{t_0}^t d\tau \ \tilde{K}(t-\tau) c(\tau)
\end{align}
where $\tilde{\omega}(k)=\omega(k)-\Omega$ with $\Omega$ the bare frequency of the resonator. However, there might be a frequency shift due to the interaction (Lamb shift); in order to compensate such shift we introduce  $c(t)=c_{\rm s}(t)e^{-i\delta\omega t}$, which leads to
\begin{align}
    &\Gamma(t)=\int_{t_0}^t d\tau \ \sum_k G_k^2 e^{-i\tilde{\omega}(k)(t-\tau)}c_{\rm s}(\tau)e^{-i\delta\omega \tau}= \notag \\& = \int_{t_0}^t d\tau \ e^{-i(\tilde{\omega}(k)-\delta\omega)(t-\tau)} e^{-i\delta\omega t}c_{\rm s}(\tau)=e^{-i\delta\omega t}\int_{t_0}^{\tau}d\tau \ K(t-\tau)c_{\rm s}(\tau).
\end{align}
Recall that we are in the rotating frame with respect to the resonator. Now, we introduce
$U(t-\tau)=\int_{t_0}^{\tau}d\tau' K(t-\tau')$ and do integration by parts as it was mentioned in the main text
\begin{align}
    &\int_{t_0}^t d\tau \ K(t-\tau)c_{\rm s}(\tau)=\int_{t_0}^td\tau \ \frac{d U(t-\tau)}{d\tau} c_{\rm s}(\tau)= \notag \\&  = U(t-\tau)c_{\rm s}(\tau) \bigg|^{t}_{t_0}-\int_{t_0}^t d\tau \ U(t-\tau)\dot{c}_{\rm s}(\tau)
\end{align}
Since $U(t-t_0)=0$, it follows that
\begin{align}
    &e^{-i\delta\omega t}\int_{t_0}^t d\tau \notag \ K(t-\tau)c_{\rm s}(\tau)=\\&c(t)\int_{t_0}^t d\tau \ K(t-\tau)-e^{-i\delta\omega t}\int_{t_0}^t d\tau \ \dot{c}_{\rm s}(\tau) \int_{t_0}^\tau d\tau' \ K(t-\tau').
\end{align}
Assuming now that $\dot{c}_{\rm s}(t)$ varies slower than the integral of the memory kernel, i.e.
\begin{align}\label{eq:KernelA}
    &e^{-i\delta\omega t}\int_{t_0}^t d\tau \notag \ K(t-\tau)c_{\rm s}(\tau)\approx \\& c(t) \int_{t_0}^td\tau \ K(t-\tau) - \dot{c}_{\rm s}(t)e^{-i\delta\omega t}\int_{t_0}^t d\tau \int_{t_0}^\tau d\tau' K(t-\tau'). 
\end{align}
Now, making use of the fact that
\begin{align}
    \dot{c}(t)= & \frac{d}{dt} \left[ c_{\rm s}(t)e^{-i\delta\omega t} \right]=\dot{c}_{\rm s}(t)e^{-i\delta\omega t}-i\delta\omega c_s(t)e^{-i\delta\omega t}= \notag \\= & \dot{c}_{\rm s}(t)e^{-i\delta\omega t}-i\delta\omega c(t),
\end{align}
we have
\begin{align}
     e^{-i\delta\omega t}\int_{t_0}^t d\tau \ K(t-\tau)c_{\rm s}(\tau)\approx c(t) K_1(t)-(\dot{c}(t)+i\delta\omega c(t))K_2(t)
\end{align}
with
\begin{align}
&K_1=\int_{t_0}^td\tau \ K(t-\tau),\\ &K_2(t)=\int_{t_0}^t d\tau \int_{t_0}^\tau d\tau' K(t-\tau'), \quad \text{and}  \\
    &K(t-\tau)=\sum_k G_k^2 e^{-i(\tilde{\omega}(k)-\delta\omega)(t-\tau)}
\end{align}
For a linear dispersion relation with $G_k\equiv G$  and $\tilde{\omega}(k)=k \Delta$, there is no Lamb shift ($\delta\omega=0$), and we can easily  compute the terms $K_1(t)$ and $K_2(t)$. Indeed, we know that $K_1(t)\approx \kappa/2$, while we identify the second term with the parameter $\mathcal{N}$, i.e.
\begin{align}\label{eq:def_a}
    \mathcal{N} =\int_{t_0}^t d\tau \int_{t_0}^\tau d\tau' K(t-\tau').
\end{align}
where $K(t-\tau')$ is in the rotating frame of the cavity. However, we should recall that the approximation in Eq.~\eqref{eq:KernelA} depends on the dynamics.


\section{Second order approximation and maximum correctable distortion}\label{app:max_distortion}

\begin{figure}[h!]
    \centering
    \includegraphics[width=.8\columnwidth]{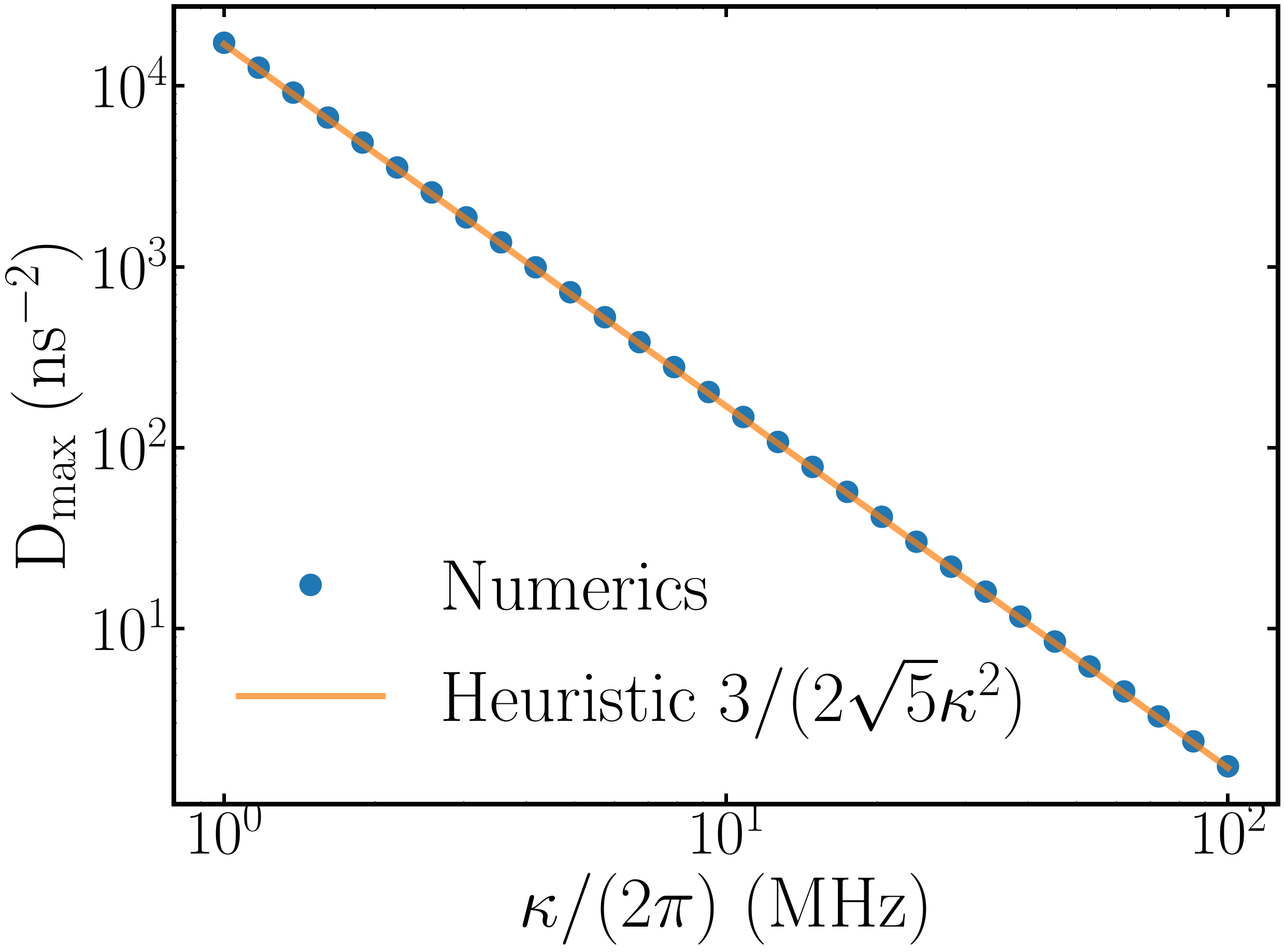}
    \caption{Maximum distortion that can be imprinted in a sech-photon using our formalism. The numerical simulations using a 3-level system (points) agree with the theoretical expression $D_{\rm max}=3/(2\sqrt{5}\kappa^2)$. }
    \label{fig:Dmax}
\end{figure}

Here we provide the details regarding the second-order approximation on the non-linear dispersion relation to obtain $g(t)$, as well as the expression for the maximum correctable distortion. We proceed as follows. Let us first take the continuum limit of infinitely many modes, so that
\begin{align}
    \xi_\text{dist}(t) &\simeq \int \mathrm{d}\omega \left(\frac{\partial \omega}{\partial k}\right)^{-1} \psi^\text{id}(k) \exp\left\{i\omega_\text{NL}(k(\omega))t_\text{AB} - i \omega t\right\}\\
    &= \int \mathrm{d}\omega \psi^\text{id}(\omega) \exp\left\{i\omega_\text{NL}(k(\omega))t_\text{AB} - i \omega t\right\}.\notag
\end{align}
From the previous equations we can compute the required pulse $g(t)$ to generate a photon with a particular distortion $D$. Indeed, let's us consider a photon $\ket{\xi(D)}=\int d\omega b_\omega^\dagger f(\omega,D) \ket{0}$ with
\begin{align}\label{eq:fomD}
    f(\omega,D)=\sqrt{\frac{\pi}{2\kappa}}{\rm sech}\left(\frac{\pi\omega}{\kappa} \right)e^{-i\omega^2 D}
\end{align}

For $D=0$, and upon a Fourier transform, we recover the photon whose time-profile is $\xi(t)=\sqrt{\kappa/4}{\rm sech}(\kappa t/2)$ . The photon is centered at the carrier frequency --taken as the origin here. For $D\neq 0$, such time-profile must be performed numerically. The aim consists in generating a photon with $D\neq0$ such that the propagation through a medium with a non-linear dispersion relation corrects this factor (we imprint a distortion with an opposite sign of the one introduced by the propagation).

From $f(\omega,D)$ we compute $\xi(t)$ and then the required pulse $g(t)$ (see App.~\ref{app:pulse_derivetion}). Yet, a physical solution demands that the qubit population remains $0\leq  |q(t)|^2\leq 1$. This physical constraint imposes a limit to the maximum distortion that can be corrected for a given $\kappa$ and a form of the injected photon. 

In order to numerically compute $D_{\rm max}(\kappa)$, we proceed as follows: For a fixed $\kappa$ we increase the distortion from $D=0$ until $0\leq |q(t)|^2\leq 1$ is not fulfilled. The last allowed value $D$ corresponds to $D_{\rm max}(\kappa)$. This is plotted in Fig.~\ref{fig:Dmax}.

In addition, we can find a heuristic argument to find the relation $D_{\rm max}(\kappa)$. The overlap between distorted and ideal photon at leading order in $D^2\kappa^4$ reads as
\begin{align}
    |z|^2=|\langle \xi(0)|\xi(D)\rangle|^2=\left| \int d\omega \ f^*(\omega,0) f(\omega,D)\right|^2= \\=\left|\int d\omega \ |f(\omega,0)|^2e^{-iD\omega^2}\right|^2 \approx 1-\frac{D^2\kappa^4}{45}+O(D^4\kappa^8).
\end{align}
Note that in the case of a travelling photon, the parameter $D$ gives account of both the dispersion relation curvature and propagation time. From the previous expression we find that $D=3\sqrt{5}(1-|z|^2)/\kappa^2$, fixing the scaling $D_{\rm max}\propto \kappa^{-2}$. By setting $|z|^2=9/10$ we find an excellent agreement with the numerical results, $D_{\rm max}=3/(2\sqrt{5}\kappa^2)$. Thus, $D_{\rm max}$ gives the maximum value for $D$ that can be imprinted in a photon, and therefore limits the amount of distortion that can be corrected.

\bibliography{paper}

\begin{thebibliography}{43}%
\makeatletter
\providecommand \@ifxundefined [1]{%
 \@ifx{#1\undefined}
}%
\providecommand \@ifnum [1]{%
 \ifnum #1\expandafter \@firstoftwo
 \else \expandafter \@secondoftwo
 \fi
}%
\providecommand \@ifx [1]{%
 \ifx #1\expandafter \@firstoftwo
 \else \expandafter \@secondoftwo
 \fi
}%
\providecommand \natexlab [1]{#1}%
\providecommand \enquote  [1]{``#1''}%
\providecommand \bibnamefont  [1]{#1}%
\providecommand \bibfnamefont [1]{#1}%
\providecommand \citenamefont [1]{#1}%
\providecommand \href@noop [0]{\@secondoftwo}%
\providecommand \href [0]{\begingroup \@sanitize@url \@href}%
\providecommand \@href[1]{\@@startlink{#1}\@@href}%
\providecommand \@@href[1]{\endgroup#1\@@endlink}%
\providecommand \@sanitize@url [0]{\catcode `\\12\catcode `\$12\catcode
  `\&12\catcode `\#12\catcode `\^12\catcode `\_12\catcode `\%12\relax}%
\providecommand \@@startlink[1]{}%
\providecommand \@@endlink[0]{}%
\providecommand \url  [0]{\begingroup\@sanitize@url \@url }%
\providecommand \@url [1]{\endgroup\@href {#1}{\urlprefix }}%
\providecommand \urlprefix  [0]{URL }%
\providecommand \Eprint [0]{\href }%
\providecommand \doibase [0]{https://doi.org/}%
\providecommand \selectlanguage [0]{\@gobble}%
\providecommand \bibinfo  [0]{\@secondoftwo}%
\providecommand \bibfield  [0]{\@secondoftwo}%
\providecommand \translation [1]{[#1]}%
\providecommand \BibitemOpen [0]{}%
\providecommand \bibitemStop [0]{}%
\providecommand \bibitemNoStop [0]{.\EOS\space}%
\providecommand \EOS [0]{\spacefactor3000\relax}%
\providecommand \BibitemShut  [1]{\csname bibitem#1\endcsname}%
\let\auto@bib@innerbib\@empty
\bibitem [{\citenamefont {Cirac}\ \emph {et~al.}(1996)\citenamefont {Cirac},
  \citenamefont {Zoller}, \citenamefont {Kimble},\ and\ \citenamefont
  {Mabuchi}}]{Cirac1996}%
  \BibitemOpen
  \bibfield  {author} {\bibinfo {author} {\bibfnamefont {J.~I.}\ \bibnamefont
  {Cirac}}, \bibinfo {author} {\bibfnamefont {P.}~\bibnamefont {Zoller}},
  \bibinfo {author} {\bibfnamefont {H.~J.}\ \bibnamefont {Kimble}},\ and\
  \bibinfo {author} {\bibfnamefont {H.}~\bibnamefont {Mabuchi}},\ }\bibfield
  {title} {\bibinfo {title} {{Quantum state transfer and entanglement
  distribution among distant nodes in a quantum network}},\ }\href
  {https://doi.org/10.1103/physrevlett.78.3221} {\bibfield  {journal} {\bibinfo
   {journal} {Phys. Rev. Lett.}\ }\textbf {\bibinfo {volume} {78}},\ \bibinfo
  {pages} {3221} (\bibinfo {year} {1996})}\BibitemShut {NoStop}%
\bibitem [{\citenamefont {Cirac}\ \emph {et~al.}(1999)\citenamefont {Cirac},
  \citenamefont {Ekert}, \citenamefont {Huelga},\ and\ \citenamefont
  {Macchiavello}}]{Cirac1999}%
  \BibitemOpen
  \bibfield  {author} {\bibinfo {author} {\bibfnamefont {J.~I.}\ \bibnamefont
  {Cirac}}, \bibinfo {author} {\bibfnamefont {A.~K.}\ \bibnamefont {Ekert}},
  \bibinfo {author} {\bibfnamefont {S.~F.}\ \bibnamefont {Huelga}},\ and\
  \bibinfo {author} {\bibfnamefont {C.}~\bibnamefont {Macchiavello}},\
  }\bibfield  {title} {\bibinfo {title} {{Distributed quantum computation over
  noisy channels}},\ }\href {https://doi.org/10.1103/PhysRevA.59.4249}
  {\bibfield  {journal} {\bibinfo  {journal} {Phys. Rev. A}\ }\textbf {\bibinfo
  {volume} {59}},\ \bibinfo {pages} {4249} (\bibinfo {year}
  {1999})}\BibitemShut {NoStop}%
\bibitem [{\citenamefont {Xiang}\ \emph {et~al.}(2017)\citenamefont {Xiang},
  \citenamefont {Zhang}, \citenamefont {Jiang},\ and\ \citenamefont
  {Rabl}}]{Xiang2017}%
  \BibitemOpen
  \bibfield  {author} {\bibinfo {author} {\bibfnamefont {Z.-L.}\ \bibnamefont
  {Xiang}}, \bibinfo {author} {\bibfnamefont {M.}~\bibnamefont {Zhang}},
  \bibinfo {author} {\bibfnamefont {L.}~\bibnamefont {Jiang}},\ and\ \bibinfo
  {author} {\bibfnamefont {P.}~\bibnamefont {Rabl}},\ }\bibfield  {title}
  {\bibinfo {title} {{Intracity Quantum Communication via Thermal Microwave
  Networks}},\ }\href {https://doi.org/10.1103/PhysRevX.7.011035} {\bibfield
  {journal} {\bibinfo  {journal} {Phys. Rev. X}\ }\textbf {\bibinfo {volume}
  {7}},\ \bibinfo {pages} {011035} (\bibinfo {year} {2017})}\BibitemShut
  {NoStop}%
\bibitem [{\citenamefont {Vogell}\ \emph {et~al.}(2017)\citenamefont {Vogell},
  \citenamefont {Vermersch}, \citenamefont {Northup}, \citenamefont {Lanyon},\
  and\ \citenamefont {Muschik}}]{Vogell2017}%
  \BibitemOpen
  \bibfield  {author} {\bibinfo {author} {\bibfnamefont {B.}~\bibnamefont
  {Vogell}}, \bibinfo {author} {\bibfnamefont {B.}~\bibnamefont {Vermersch}},
  \bibinfo {author} {\bibfnamefont {T.~E.}\ \bibnamefont {Northup}}, \bibinfo
  {author} {\bibfnamefont {B.~P.}\ \bibnamefont {Lanyon}},\ and\ \bibinfo
  {author} {\bibfnamefont {C.~A.}\ \bibnamefont {Muschik}},\ }\bibfield
  {title} {\bibinfo {title} {{Deterministic quantum state transfer between
  remote qubits in cavities}},\ }\href
  {https://doi.org/10.1088/2058-9565/aa868b} {\bibfield  {journal} {\bibinfo
  {journal} {Quantum Sci. Technol.}\ }\textbf {\bibinfo {volume} {2}},\
  \bibinfo {pages} {045003} (\bibinfo {year} {2017})}\BibitemShut {NoStop}%
\bibitem [{\citenamefont {Yao}\ \emph {et~al.}(2011)\citenamefont {Yao},
  \citenamefont {Jiang}, \citenamefont {Gorshkov}, \citenamefont {Gong},
  \citenamefont {Zhai}, \citenamefont {Duan},\ and\ \citenamefont
  {Lukin}}]{Yao2011}%
  \BibitemOpen
  \bibfield  {author} {\bibinfo {author} {\bibfnamefont {N.~Y.}\ \bibnamefont
  {Yao}}, \bibinfo {author} {\bibfnamefont {L.}~\bibnamefont {Jiang}}, \bibinfo
  {author} {\bibfnamefont {A.~V.}\ \bibnamefont {Gorshkov}}, \bibinfo {author}
  {\bibfnamefont {Z.-X.}\ \bibnamefont {Gong}}, \bibinfo {author}
  {\bibfnamefont {A.}~\bibnamefont {Zhai}}, \bibinfo {author} {\bibfnamefont
  {L.-M.}\ \bibnamefont {Duan}},\ and\ \bibinfo {author} {\bibfnamefont
  {M.~D.}\ \bibnamefont {Lukin}},\ }\bibfield  {title} {\bibinfo {title}
  {Robust quantum state transfer in random unpolarized spin chains},\ }\href
  {https://doi.org/10.1103/PhysRevLett.106.040505} {\bibfield  {journal}
  {\bibinfo  {journal} {Phys. Rev. Lett.}\ }\textbf {\bibinfo {volume} {106}},\
  \bibinfo {pages} {040505} (\bibinfo {year} {2011})}\BibitemShut {NoStop}%
\bibitem [{\citenamefont {Stannigel}\ \emph {et~al.}(2010)\citenamefont
  {Stannigel}, \citenamefont {Rabl}, \citenamefont {S\o{}rensen}, \citenamefont
  {Zoller},\ and\ \citenamefont {Lukin}}]{Stannigel2010}%
  \BibitemOpen
  \bibfield  {author} {\bibinfo {author} {\bibfnamefont {K.}~\bibnamefont
  {Stannigel}}, \bibinfo {author} {\bibfnamefont {P.}~\bibnamefont {Rabl}},
  \bibinfo {author} {\bibfnamefont {A.~S.}\ \bibnamefont {S\o{}rensen}},
  \bibinfo {author} {\bibfnamefont {P.}~\bibnamefont {Zoller}},\ and\ \bibinfo
  {author} {\bibfnamefont {M.~D.}\ \bibnamefont {Lukin}},\ }\bibfield  {title}
  {\bibinfo {title} {Optomechanical transducers for long-distance quantum
  communication},\ }\href {https://doi.org/10.1103/PhysRevLett.105.220501}
  {\bibfield  {journal} {\bibinfo  {journal} {Phys. Rev. Lett.}\ }\textbf
  {\bibinfo {volume} {105}},\ \bibinfo {pages} {220501} (\bibinfo {year}
  {2010})}\BibitemShut {NoStop}%
\bibitem [{\citenamefont {Stannigel}\ \emph {et~al.}(2011)\citenamefont
  {Stannigel}, \citenamefont {Rabl}, \citenamefont {S\o{}rensen}, \citenamefont
  {Lukin},\ and\ \citenamefont {Zoller}}]{Stannigel2011}%
  \BibitemOpen
  \bibfield  {author} {\bibinfo {author} {\bibfnamefont {K.}~\bibnamefont
  {Stannigel}}, \bibinfo {author} {\bibfnamefont {P.}~\bibnamefont {Rabl}},
  \bibinfo {author} {\bibfnamefont {A.~S.}\ \bibnamefont {S\o{}rensen}},
  \bibinfo {author} {\bibfnamefont {M.~D.}\ \bibnamefont {Lukin}},\ and\
  \bibinfo {author} {\bibfnamefont {P.}~\bibnamefont {Zoller}},\ }\bibfield
  {title} {\bibinfo {title} {Optomechanical transducers for quantum-information
  processing},\ }\href {https://doi.org/10.1103/PhysRevA.84.042341} {\bibfield
  {journal} {\bibinfo  {journal} {Phys. Rev. A}\ }\textbf {\bibinfo {volume}
  {84}},\ \bibinfo {pages} {042341} (\bibinfo {year} {2011})}\BibitemShut
  {NoStop}%
\bibitem [{\citenamefont {Wang}\ and\ \citenamefont {Clerk}(2011)}]{Wang2011}%
  \BibitemOpen
  \bibfield  {author} {\bibinfo {author} {\bibfnamefont {Y.-D.}\ \bibnamefont
  {Wang}}\ and\ \bibinfo {author} {\bibfnamefont {A.~A.}\ \bibnamefont
  {Clerk}},\ }\bibfield  {title} {\bibinfo {title} {{Using interference for
  high fidelity quantum state transfer in optomechanics}},\ }\href
  {https://doi.org/10.1103/physrevlett.108.153603} {\bibfield  {journal}
  {\bibinfo  {journal} {Phys. Rev. Lett.}\ }\textbf {\bibinfo {volume} {108}}
  (\bibinfo {year} {2011})}\BibitemShut {NoStop}%
\bibitem [{\citenamefont {Lemonde}\ \emph {et~al.}(2018)\citenamefont
  {Lemonde}, \citenamefont {Meesala}, \citenamefont {Sipahigil}, \citenamefont
  {Schuetz}, \citenamefont {Lukin}, \citenamefont {Loncar},\ and\ \citenamefont
  {Rabl}}]{Lemonde2018}%
  \BibitemOpen
  \bibfield  {author} {\bibinfo {author} {\bibfnamefont {M.-A.}\ \bibnamefont
  {Lemonde}}, \bibinfo {author} {\bibfnamefont {S.}~\bibnamefont {Meesala}},
  \bibinfo {author} {\bibfnamefont {A.}~\bibnamefont {Sipahigil}}, \bibinfo
  {author} {\bibfnamefont {M.~J.~A.}\ \bibnamefont {Schuetz}}, \bibinfo
  {author} {\bibfnamefont {M.~D.}\ \bibnamefont {Lukin}}, \bibinfo {author}
  {\bibfnamefont {M.}~\bibnamefont {Loncar}},\ and\ \bibinfo {author}
  {\bibfnamefont {P.}~\bibnamefont {Rabl}},\ }\bibfield  {title} {\bibinfo
  {title} {{Phonon Networks with Silicon-Vacancy Centers in Diamond
  Waveguides}},\ }\href {https://doi.org/10.1103/PhysRevLett.120.213603}
  {\bibfield  {journal} {\bibinfo  {journal} {Phys. Rev. Lett.}\ }\textbf
  {\bibinfo {volume} {120}},\ \bibinfo {pages} {213603} (\bibinfo {year}
  {2018})}\BibitemShut {NoStop}%
\bibitem [{\citenamefont {Calaj\'o}\ \emph {et~al.}(2019)\citenamefont
  {Calaj\'o}, \citenamefont {Schuetz}, \citenamefont {Pichler}, \citenamefont
  {Lukin}, \citenamefont {Schneeweiss}, \citenamefont {Volz},\ and\
  \citenamefont {Rabl}}]{Calajo2019}%
  \BibitemOpen
  \bibfield  {author} {\bibinfo {author} {\bibfnamefont {G.}~\bibnamefont
  {Calaj\'o}}, \bibinfo {author} {\bibfnamefont {M.~J.~A.}\ \bibnamefont
  {Schuetz}}, \bibinfo {author} {\bibfnamefont {H.}~\bibnamefont {Pichler}},
  \bibinfo {author} {\bibfnamefont {M.~D.}\ \bibnamefont {Lukin}}, \bibinfo
  {author} {\bibfnamefont {P.}~\bibnamefont {Schneeweiss}}, \bibinfo {author}
  {\bibfnamefont {J.}~\bibnamefont {Volz}},\ and\ \bibinfo {author}
  {\bibfnamefont {P.}~\bibnamefont {Rabl}},\ }\bibfield  {title} {\bibinfo
  {title} {Quantum acousto-optic control of light-matter interactions in
  nanophotonic networks},\ }\href {https://doi.org/10.1103/PhysRevA.99.053852}
  {\bibfield  {journal} {\bibinfo  {journal} {Phys. Rev. A}\ }\textbf {\bibinfo
  {volume} {99}},\ \bibinfo {pages} {053852} (\bibinfo {year}
  {2019})}\BibitemShut {NoStop}%
\bibitem [{\citenamefont {Vermersch}\ \emph {et~al.}(2017)\citenamefont
  {Vermersch}, \citenamefont {Guimond}, \citenamefont {Pichler},\ and\
  \citenamefont {Zoller}}]{Vermersch2016}%
  \BibitemOpen
  \bibfield  {author} {\bibinfo {author} {\bibfnamefont {B.}~\bibnamefont
  {Vermersch}}, \bibinfo {author} {\bibfnamefont {P.-O.}\ \bibnamefont
  {Guimond}}, \bibinfo {author} {\bibfnamefont {H.}~\bibnamefont {Pichler}},\
  and\ \bibinfo {author} {\bibfnamefont {P.}~\bibnamefont {Zoller}},\
  }\bibfield  {title} {\bibinfo {title} {Quantum state transfer via noisy
  photonic and phononic waveguides},\ }\href
  {https://doi.org/10.1103/PhysRevLett.118.133601} {\bibfield  {journal}
  {\bibinfo  {journal} {Phys. Rev. Lett.}\ }\textbf {\bibinfo {volume} {118}},\
  \bibinfo {pages} {133601} (\bibinfo {year} {2017})}\BibitemShut {NoStop}%
\bibitem [{\citenamefont {Bennett}\ \emph {et~al.}(1993)\citenamefont
  {Bennett}, \citenamefont {Brassard}, \citenamefont {Cr\'epeau}, \citenamefont
  {Jozsa}, \citenamefont {Peres},\ and\ \citenamefont {Wootters}}]{Bennet1993}%
  \BibitemOpen
  \bibfield  {author} {\bibinfo {author} {\bibfnamefont {C.~H.}\ \bibnamefont
  {Bennett}}, \bibinfo {author} {\bibfnamefont {G.}~\bibnamefont {Brassard}},
  \bibinfo {author} {\bibfnamefont {C.}~\bibnamefont {Cr\'epeau}}, \bibinfo
  {author} {\bibfnamefont {R.}~\bibnamefont {Jozsa}}, \bibinfo {author}
  {\bibfnamefont {A.}~\bibnamefont {Peres}},\ and\ \bibinfo {author}
  {\bibfnamefont {W.~K.}\ \bibnamefont {Wootters}},\ }\bibfield  {title}
  {\bibinfo {title} {Teleporting an unknown quantum state via dual classical
  and einstein-podolsky-rosen channels},\ }\href
  {https://doi.org/10.1103/PhysRevLett.70.1895} {\bibfield  {journal} {\bibinfo
   {journal} {Phys. Rev. Lett.}\ }\textbf {\bibinfo {volume} {70}},\ \bibinfo
  {pages} {1895} (\bibinfo {year} {1993})}\BibitemShut {NoStop}%
\bibitem [{\citenamefont {Awschalom}\ \emph {et~al.}(2021)\citenamefont
  {Awschalom}, \citenamefont {Berggren}, \citenamefont {Bernien}, \citenamefont
  {Bhave}, \citenamefont {Carr}, \citenamefont {Davids}, \citenamefont
  {Economou}, \citenamefont {Englund}, \citenamefont {Faraon}, \citenamefont
  {Fejer}, \citenamefont {Guha}, \citenamefont {Gustafsson}, \citenamefont
  {Hu}, \citenamefont {Jiang}, \citenamefont {Kim}, \citenamefont {Korzh},
  \citenamefont {Kumar}, \citenamefont {Kwiat}, \citenamefont
  {Lon\ifmmode~\check{c}\else \v{c}\fi{}ar}, \citenamefont {Lukin},
  \citenamefont {Miller}, \citenamefont {Monroe}, \citenamefont {Nam},
  \citenamefont {Narang}, \citenamefont {Orcutt}, \citenamefont {Raymer},
  \citenamefont {Safavi-Naeini}, \citenamefont {Spiropulu}, \citenamefont
  {Srinivasan}, \citenamefont {Sun}, \citenamefont {Vu\ifmmode \check{c}\else
  \v{c}\fi{}kovi\ifmmode~\acute{c}\else \'{c}\fi{}}, \citenamefont {Waks},
  \citenamefont {Walsworth}, \citenamefont {Weiner},\ and\ \citenamefont
  {Zhang}}]{Awschalom2021}%
  \BibitemOpen
  \bibfield  {author} {\bibinfo {author} {\bibfnamefont {D.}~\bibnamefont
  {Awschalom}}, \bibinfo {author} {\bibfnamefont {K.~K.}\ \bibnamefont
  {Berggren}}, \bibinfo {author} {\bibfnamefont {H.}~\bibnamefont {Bernien}},
  \bibinfo {author} {\bibfnamefont {S.}~\bibnamefont {Bhave}}, \bibinfo
  {author} {\bibfnamefont {L.~D.}\ \bibnamefont {Carr}}, \bibinfo {author}
  {\bibfnamefont {P.}~\bibnamefont {Davids}}, \bibinfo {author} {\bibfnamefont
  {S.~E.}\ \bibnamefont {Economou}}, \bibinfo {author} {\bibfnamefont
  {D.}~\bibnamefont {Englund}}, \bibinfo {author} {\bibfnamefont
  {A.}~\bibnamefont {Faraon}}, \bibinfo {author} {\bibfnamefont
  {M.}~\bibnamefont {Fejer}}, \bibinfo {author} {\bibfnamefont
  {S.}~\bibnamefont {Guha}}, \bibinfo {author} {\bibfnamefont {M.~V.}\
  \bibnamefont {Gustafsson}}, \bibinfo {author} {\bibfnamefont
  {E.}~\bibnamefont {Hu}}, \bibinfo {author} {\bibfnamefont {L.}~\bibnamefont
  {Jiang}}, \bibinfo {author} {\bibfnamefont {J.}~\bibnamefont {Kim}}, \bibinfo
  {author} {\bibfnamefont {B.}~\bibnamefont {Korzh}}, \bibinfo {author}
  {\bibfnamefont {P.}~\bibnamefont {Kumar}}, \bibinfo {author} {\bibfnamefont
  {P.~G.}\ \bibnamefont {Kwiat}}, \bibinfo {author} {\bibfnamefont
  {M.}~\bibnamefont {Lon\ifmmode~\check{c}\else \v{c}\fi{}ar}}, \bibinfo
  {author} {\bibfnamefont {M.~D.}\ \bibnamefont {Lukin}}, \bibinfo {author}
  {\bibfnamefont {D.~A.}\ \bibnamefont {Miller}}, \bibinfo {author}
  {\bibfnamefont {C.}~\bibnamefont {Monroe}}, \bibinfo {author} {\bibfnamefont
  {S.~W.}\ \bibnamefont {Nam}}, \bibinfo {author} {\bibfnamefont
  {P.}~\bibnamefont {Narang}}, \bibinfo {author} {\bibfnamefont {J.~S.}\
  \bibnamefont {Orcutt}}, \bibinfo {author} {\bibfnamefont {M.~G.}\
  \bibnamefont {Raymer}}, \bibinfo {author} {\bibfnamefont {A.~H.}\
  \bibnamefont {Safavi-Naeini}}, \bibinfo {author} {\bibfnamefont
  {M.}~\bibnamefont {Spiropulu}}, \bibinfo {author} {\bibfnamefont
  {K.}~\bibnamefont {Srinivasan}}, \bibinfo {author} {\bibfnamefont
  {S.}~\bibnamefont {Sun}}, \bibinfo {author} {\bibfnamefont {J.}~\bibnamefont
  {Vu\ifmmode \check{c}\else \v{c}\fi{}kovi\ifmmode~\acute{c}\else
  \'{c}\fi{}}}, \bibinfo {author} {\bibfnamefont {E.}~\bibnamefont {Waks}},
  \bibinfo {author} {\bibfnamefont {R.}~\bibnamefont {Walsworth}}, \bibinfo
  {author} {\bibfnamefont {A.~M.}\ \bibnamefont {Weiner}},\ and\ \bibinfo
  {author} {\bibfnamefont {Z.}~\bibnamefont {Zhang}},\ }\bibfield  {title}
  {\bibinfo {title} {Development of quantum interconnects (quics) for
  next-generation information technologies},\ }\href
  {https://doi.org/10.1103/PRXQuantum.2.017002} {\bibfield  {journal} {\bibinfo
   {journal} {PRX Quantum}\ }\textbf {\bibinfo {volume} {2}},\ \bibinfo {pages}
  {017002} (\bibinfo {year} {2021})}\BibitemShut {NoStop}%
\bibitem [{\citenamefont {Kimble}(2008)}]{Kimble2008}%
  \BibitemOpen
  \bibfield  {author} {\bibinfo {author} {\bibfnamefont {H.~J.}\ \bibnamefont
  {Kimble}},\ }\bibfield  {title} {\bibinfo {title} {{The quantum internet}},\
  }\href {https://doi.org/10.1038/nature07127} {\bibfield  {journal} {\bibinfo
  {journal} {Nature}\ }\textbf {\bibinfo {volume} {453}},\ \bibinfo {pages}
  {1023} (\bibinfo {year} {2008})}\BibitemShut {NoStop}%
\bibitem [{\citenamefont {Wehner}\ \emph {et~al.}(2018)\citenamefont {Wehner},
  \citenamefont {Elkouss},\ and\ \citenamefont {Hanson}}]{Wehner2018}%
  \BibitemOpen
  \bibfield  {author} {\bibinfo {author} {\bibfnamefont {S.}~\bibnamefont
  {Wehner}}, \bibinfo {author} {\bibfnamefont {D.}~\bibnamefont {Elkouss}},\
  and\ \bibinfo {author} {\bibfnamefont {R.}~\bibnamefont {Hanson}},\
  }\bibfield  {title} {\bibinfo {title} {{Quantum internet: A vision for the
  road ahead}},\ }\href {https://doi.org/10.1126/science.aam9288} {\bibfield
  {journal} {\bibinfo  {journal} {Science}\ }\textbf {\bibinfo {volume}
  {362}},\ \bibinfo {pages} {6412} (\bibinfo {year} {2018})}\BibitemShut
  {NoStop}%
\bibitem [{\citenamefont {Cacciapuoti}\ \emph {et~al.}(2020)\citenamefont
  {Cacciapuoti}, \citenamefont {Caleffi}, \citenamefont {Tafuri}, \citenamefont
  {Cataliotti}, \citenamefont {Gherardini},\ and\ \citenamefont
  {Bianchi}}]{Cacciapuoti2020}%
  \BibitemOpen
  \bibfield  {author} {\bibinfo {author} {\bibfnamefont {A.~S.}\ \bibnamefont
  {Cacciapuoti}}, \bibinfo {author} {\bibfnamefont {M.}~\bibnamefont
  {Caleffi}}, \bibinfo {author} {\bibfnamefont {F.}~\bibnamefont {Tafuri}},
  \bibinfo {author} {\bibfnamefont {F.~S.}\ \bibnamefont {Cataliotti}},
  \bibinfo {author} {\bibfnamefont {S.}~\bibnamefont {Gherardini}},\ and\
  \bibinfo {author} {\bibfnamefont {G.}~\bibnamefont {Bianchi}},\ }\bibfield
  {title} {\bibinfo {title} {Quantum internet: Networking challenges in
  distributed quantum computing},\ }\href
  {https://doi.org/10.1109/MNET.001.1900092} {\bibfield  {journal} {\bibinfo
  {journal} {IEEE Network}\ }\textbf {\bibinfo {volume} {34}},\ \bibinfo
  {pages} {137} (\bibinfo {year} {2020})}\BibitemShut {NoStop}%
\bibitem [{\citenamefont {Magnard}\ \emph {et~al.}(2020)\citenamefont
  {Magnard}, \citenamefont {Storz}, \citenamefont {Kurpiers}, \citenamefont
  {Sch{\"{a}}r}, \citenamefont {Marxer}, \citenamefont {L{\"{u}}tolf},
  \citenamefont {Walter}, \citenamefont {Besse}, \citenamefont {Gabureac},
  \citenamefont {Reuer}, \citenamefont {Akin}, \citenamefont {Royer},
  \citenamefont {Blais},\ and\ \citenamefont {Wallraff}}]{Magnard2020}%
  \BibitemOpen
  \bibfield  {author} {\bibinfo {author} {\bibfnamefont {P.}~\bibnamefont
  {Magnard}}, \bibinfo {author} {\bibfnamefont {S.}~\bibnamefont {Storz}},
  \bibinfo {author} {\bibfnamefont {P.}~\bibnamefont {Kurpiers}}, \bibinfo
  {author} {\bibfnamefont {J.}~\bibnamefont {Sch{\"{a}}r}}, \bibinfo {author}
  {\bibfnamefont {F.}~\bibnamefont {Marxer}}, \bibinfo {author} {\bibfnamefont
  {J.}~\bibnamefont {L{\"{u}}tolf}}, \bibinfo {author} {\bibfnamefont
  {T.}~\bibnamefont {Walter}}, \bibinfo {author} {\bibfnamefont {J.-C.}\
  \bibnamefont {Besse}}, \bibinfo {author} {\bibfnamefont {M.}~\bibnamefont
  {Gabureac}}, \bibinfo {author} {\bibfnamefont {K.}~\bibnamefont {Reuer}},
  \bibinfo {author} {\bibfnamefont {A.}~\bibnamefont {Akin}}, \bibinfo {author}
  {\bibfnamefont {B.}~\bibnamefont {Royer}}, \bibinfo {author} {\bibfnamefont
  {A.}~\bibnamefont {Blais}},\ and\ \bibinfo {author} {\bibfnamefont
  {A.}~\bibnamefont {Wallraff}},\ }\bibfield  {title} {\bibinfo {title}
  {{Microwave Quantum Link between Superconducting Circuits Housed in Spatially
  Separated Cryogenic Systems}},\ }\href
  {https://doi.org/10.1103/PhysRevLett.125.260502} {\bibfield  {journal}
  {\bibinfo  {journal} {Phys. Rev. Lett.}\ }\textbf {\bibinfo {volume} {125}},\
  \bibinfo {pages} {260502} (\bibinfo {year} {2020})}\BibitemShut {NoStop}%
\bibitem [{\citenamefont {Kurpiers}\ \emph {et~al.}(2017)\citenamefont
  {Kurpiers}, \citenamefont {Magnard}, \citenamefont {Walter}, \citenamefont
  {Royer}, \citenamefont {Pechal}, \citenamefont {Heinsoo}, \citenamefont
  {Salath{\'{e}}}, \citenamefont {Akin}, \citenamefont {Storz}, \citenamefont
  {Besse}, \citenamefont {Gasparinetti}, \citenamefont {Blais},\ and\
  \citenamefont {Wallraff}}]{Kurpiers2017}%
  \BibitemOpen
  \bibfield  {author} {\bibinfo {author} {\bibfnamefont {P.}~\bibnamefont
  {Kurpiers}}, \bibinfo {author} {\bibfnamefont {P.}~\bibnamefont {Magnard}},
  \bibinfo {author} {\bibfnamefont {T.}~\bibnamefont {Walter}}, \bibinfo
  {author} {\bibfnamefont {B.}~\bibnamefont {Royer}}, \bibinfo {author}
  {\bibfnamefont {M.}~\bibnamefont {Pechal}}, \bibinfo {author} {\bibfnamefont
  {J.}~\bibnamefont {Heinsoo}}, \bibinfo {author} {\bibfnamefont
  {Y.}~\bibnamefont {Salath{\'{e}}}}, \bibinfo {author} {\bibfnamefont
  {A.}~\bibnamefont {Akin}}, \bibinfo {author} {\bibfnamefont {S.}~\bibnamefont
  {Storz}}, \bibinfo {author} {\bibfnamefont {J.-C.}\ \bibnamefont {Besse}},
  \bibinfo {author} {\bibfnamefont {S.}~\bibnamefont {Gasparinetti}}, \bibinfo
  {author} {\bibfnamefont {A.}~\bibnamefont {Blais}},\ and\ \bibinfo {author}
  {\bibfnamefont {A.}~\bibnamefont {Wallraff}},\ }\bibfield  {title} {\bibinfo
  {title} {{Deterministic Quantum State Transfer and Generation of Remote
  Entanglement using Microwave Photons}},\ }\href
  {https://doi.org/10.1038/s41586-018-0195-y} {\bibfield  {journal} {\bibinfo
  {journal} {Nature}\ }\textbf {\bibinfo {volume} {558}},\ \bibinfo {pages}
  {264} (\bibinfo {year} {2017})}\BibitemShut {NoStop}%
\bibitem [{\citenamefont {Leung}\ \emph {et~al.}(2019)\citenamefont {Leung},
  \citenamefont {Lu}, \citenamefont {Chakram}, \citenamefont {Naik},
  \citenamefont {Earnest}, \citenamefont {Ma}, \citenamefont {Jacobs},
  \citenamefont {Cleland},\ and\ \citenamefont {Schuster}}]{Leung2019}%
  \BibitemOpen
  \bibfield  {author} {\bibinfo {author} {\bibfnamefont {N.}~\bibnamefont
  {Leung}}, \bibinfo {author} {\bibfnamefont {Y.}~\bibnamefont {Lu}}, \bibinfo
  {author} {\bibfnamefont {S.}~\bibnamefont {Chakram}}, \bibinfo {author}
  {\bibfnamefont {R.~K.}\ \bibnamefont {Naik}}, \bibinfo {author}
  {\bibfnamefont {N.}~\bibnamefont {Earnest}}, \bibinfo {author} {\bibfnamefont
  {R.}~\bibnamefont {Ma}}, \bibinfo {author} {\bibfnamefont {K.}~\bibnamefont
  {Jacobs}}, \bibinfo {author} {\bibfnamefont {A.~N.}\ \bibnamefont
  {Cleland}},\ and\ \bibinfo {author} {\bibfnamefont {D.~I.}\ \bibnamefont
  {Schuster}},\ }\bibfield  {title} {\bibinfo {title} {{Deterministic
  bidirectional communication and remote entanglement generation between
  superconducting qubits}},\ }\href {https://doi.org/10.1038/s41534-019-0128-0}
  {\bibfield  {journal} {\bibinfo  {journal} {npj Quantum Information}\
  }\textbf {\bibinfo {volume} {5}},\ \bibinfo {pages} {1} (\bibinfo {year}
  {2019})}\BibitemShut {NoStop}%
\bibitem [{\citenamefont {Chang}\ \emph {et~al.}(2020)\citenamefont {Chang},
  \citenamefont {Zhong}, \citenamefont {Bienfait}, \citenamefont {Chou},
  \citenamefont {Conner}, \citenamefont {Dumur}, \citenamefont {Grebel},
  \citenamefont {Peairs}, \citenamefont {Povey}, \citenamefont {Satzinger},\
  and\ \citenamefont {Cleland}}]{Chang2020}%
  \BibitemOpen
  \bibfield  {author} {\bibinfo {author} {\bibfnamefont {H.-S.}\ \bibnamefont
  {Chang}}, \bibinfo {author} {\bibfnamefont {Y.~P.}\ \bibnamefont {Zhong}},
  \bibinfo {author} {\bibfnamefont {A.}~\bibnamefont {Bienfait}}, \bibinfo
  {author} {\bibfnamefont {M.-H.}\ \bibnamefont {Chou}}, \bibinfo {author}
  {\bibfnamefont {C.~R.}\ \bibnamefont {Conner}}, \bibinfo {author}
  {\bibfnamefont {E.}~\bibnamefont {Dumur}}, \bibinfo {author} {\bibfnamefont
  {J.}~\bibnamefont {Grebel}}, \bibinfo {author} {\bibfnamefont {G.~A.}\
  \bibnamefont {Peairs}}, \bibinfo {author} {\bibfnamefont {R.~G.}\
  \bibnamefont {Povey}}, \bibinfo {author} {\bibfnamefont {K.~J.}\ \bibnamefont
  {Satzinger}},\ and\ \bibinfo {author} {\bibfnamefont {A.~N.}\ \bibnamefont
  {Cleland}},\ }\bibfield  {title} {\bibinfo {title} {Remote entanglement via
  adiabatic passage using a tunably dissipative quantum communication system},\
  }\href {https://doi.org/10.1103/PhysRevLett.124.240502} {\bibfield  {journal}
  {\bibinfo  {journal} {Phys. Rev. Lett.}\ }\textbf {\bibinfo {volume} {124}},\
  \bibinfo {pages} {240502} (\bibinfo {year} {2020})}\BibitemShut {NoStop}%
\bibitem [{\citenamefont {Ritter}\ \emph {et~al.}(2012)\citenamefont {Ritter},
  \citenamefont {N{\"{o}}lleke}, \citenamefont {Hahn}, \citenamefont
  {Reiserer}, \citenamefont {Neuzner}, \citenamefont {Uphoff}, \citenamefont
  {M{\"{u}}cke}, \citenamefont {Figueroa}, \citenamefont {Bochmann},\ and\
  \citenamefont {Rempe}}]{Ritter2012}%
  \BibitemOpen
  \bibfield  {author} {\bibinfo {author} {\bibfnamefont {S.}~\bibnamefont
  {Ritter}}, \bibinfo {author} {\bibfnamefont {C.}~\bibnamefont
  {N{\"{o}}lleke}}, \bibinfo {author} {\bibfnamefont {C.}~\bibnamefont {Hahn}},
  \bibinfo {author} {\bibfnamefont {A.}~\bibnamefont {Reiserer}}, \bibinfo
  {author} {\bibfnamefont {A.}~\bibnamefont {Neuzner}}, \bibinfo {author}
  {\bibfnamefont {M.}~\bibnamefont {Uphoff}}, \bibinfo {author} {\bibfnamefont
  {M.}~\bibnamefont {M{\"{u}}cke}}, \bibinfo {author} {\bibfnamefont
  {E.}~\bibnamefont {Figueroa}}, \bibinfo {author} {\bibfnamefont
  {J.}~\bibnamefont {Bochmann}},\ and\ \bibinfo {author} {\bibfnamefont
  {G.}~\bibnamefont {Rempe}},\ }\bibfield  {title} {\bibinfo {title} {{An
  Elementary Quantum Network of Single Atoms in Optical Cavities}},\ }\href
  {https://doi.org/10.1038/nature11023} {\bibfield  {journal} {\bibinfo
  {journal} {Nature}\ }\textbf {\bibinfo {volume} {484}},\ \bibinfo {pages}
  {195} (\bibinfo {year} {2012})}\BibitemShut {NoStop}%
\bibitem [{\citenamefont {Bienfait}\ \emph {et~al.}(2019)\citenamefont
  {Bienfait}, \citenamefont {Satzinger}, \citenamefont {Zhong}, \citenamefont
  {Chang}, \citenamefont {Chou}, \citenamefont {Conner}, \citenamefont {Dumur},
  \citenamefont {Grebel}, \citenamefont {Peairs}, \citenamefont {Povey},\ and\
  \citenamefont {Cleland}}]{Bienfait2019}%
  \BibitemOpen
  \bibfield  {author} {\bibinfo {author} {\bibfnamefont {A.}~\bibnamefont
  {Bienfait}}, \bibinfo {author} {\bibfnamefont {K.~J.}\ \bibnamefont
  {Satzinger}}, \bibinfo {author} {\bibfnamefont {Y.~P.}\ \bibnamefont
  {Zhong}}, \bibinfo {author} {\bibfnamefont {H.~S.}\ \bibnamefont {Chang}},
  \bibinfo {author} {\bibfnamefont {M.~H.}\ \bibnamefont {Chou}}, \bibinfo
  {author} {\bibfnamefont {C.~R.}\ \bibnamefont {Conner}}, \bibinfo {author}
  {\bibnamefont {Dumur}}, \bibinfo {author} {\bibfnamefont {J.}~\bibnamefont
  {Grebel}}, \bibinfo {author} {\bibfnamefont {G.~A.}\ \bibnamefont {Peairs}},
  \bibinfo {author} {\bibfnamefont {R.~G.}\ \bibnamefont {Povey}},\ and\
  \bibinfo {author} {\bibfnamefont {A.~N.}\ \bibnamefont {Cleland}},\
  }\bibfield  {title} {\bibinfo {title} {{Phonon-mediated quantum state
  transfer and remote qubit entanglement}},\ }\href
  {https://doi.org/10.1126/SCIENCE.AAW8415} {\bibfield  {journal} {\bibinfo
  {journal} {Science}\ }\textbf {\bibinfo {volume} {364}},\ \bibinfo {pages}
  {368} (\bibinfo {year} {2019})}\BibitemShut {NoStop}%
\bibitem [{\citenamefont {Pe\~nas}\ \emph {et~al.}(2022)\citenamefont
  {Pe\~nas}, \citenamefont {Puebla}, \citenamefont {Ramos}, \citenamefont
  {Rabl},\ and\ \citenamefont {Garc\'{\i}a-Ripoll}}]{Penas2022}%
  \BibitemOpen
  \bibfield  {author} {\bibinfo {author} {\bibfnamefont {G.~F.}\ \bibnamefont
  {Pe\~nas}}, \bibinfo {author} {\bibfnamefont {R.}~\bibnamefont {Puebla}},
  \bibinfo {author} {\bibfnamefont {T.}~\bibnamefont {Ramos}}, \bibinfo
  {author} {\bibfnamefont {P.}~\bibnamefont {Rabl}},\ and\ \bibinfo {author}
  {\bibfnamefont {J.~J.}\ \bibnamefont {Garc\'{\i}a-Ripoll}},\ }\bibfield
  {title} {\bibinfo {title} {Universal deterministic quantum operations in
  microwave quantum links},\ }\href
  {https://doi.org/10.1103/PhysRevApplied.17.054038} {\bibfield  {journal}
  {\bibinfo  {journal} {Phys. Rev. Applied}\ }\textbf {\bibinfo {volume}
  {17}},\ \bibinfo {pages} {054038} (\bibinfo {year} {2022})}\BibitemShut
  {NoStop}%
\bibitem [{\citenamefont {Yu}\ \emph {et~al.}(2019)\citenamefont {Yu},
  \citenamefont {Muniz}, \citenamefont {Hung},\ and\ \citenamefont
  {Kimble}}]{Yu19}%
  \BibitemOpen
  \bibfield  {author} {\bibinfo {author} {\bibfnamefont {S.-P.}\ \bibnamefont
  {Yu}}, \bibinfo {author} {\bibfnamefont {J.~A.}\ \bibnamefont {Muniz}},
  \bibinfo {author} {\bibfnamefont {C.-L.}\ \bibnamefont {Hung}},\ and\
  \bibinfo {author} {\bibfnamefont {H.~J.}\ \bibnamefont {Kimble}},\ }\bibfield
   {title} {\bibinfo {title} {Two-dimensional photonic crystals for engineering
  atom–light interactions},\ }\href {https://doi.org/10.1073/pnas.1822110116}
  {\bibfield  {journal} {\bibinfo  {journal} {PNAS}\ }\textbf {\bibinfo
  {volume} {116}},\ \bibinfo {pages} {12743} (\bibinfo {year}
  {2019})}\BibitemShut {NoStop}%
\bibitem [{\citenamefont {Butt}\ \emph {et~al.}(2021)\citenamefont {Butt},
  \citenamefont {Khonina},\ and\ \citenamefont {Kazanskiy}}]{Butt21}%
  \BibitemOpen
  \bibfield  {author} {\bibinfo {author} {\bibfnamefont {M.}~\bibnamefont
  {Butt}}, \bibinfo {author} {\bibfnamefont {S.}~\bibnamefont {Khonina}},\ and\
  \bibinfo {author} {\bibfnamefont {N.}~\bibnamefont {Kazanskiy}},\ }\bibfield
  {title} {\bibinfo {title} {Recent advances in photonic crystal optical
  devices: A review},\ }\href
  {https://doi.org/https://doi.org/10.1016/j.optlastec.2021.107265} {\bibfield
  {journal} {\bibinfo  {journal} {Optics \& Laser Technology}\ }\textbf
  {\bibinfo {volume} {142}},\ \bibinfo {pages} {107265} (\bibinfo {year}
  {2021})}\BibitemShut {NoStop}%
\bibitem [{\citenamefont {Ramos}\ \emph {et~al.}(2016)\citenamefont {Ramos},
  \citenamefont {Vermersch}, \citenamefont {Hauke}, \citenamefont {Pichler},\
  and\ \citenamefont {Zoller}}]{ramos2016non}%
  \BibitemOpen
  \bibfield  {author} {\bibinfo {author} {\bibfnamefont {T.}~\bibnamefont
  {Ramos}}, \bibinfo {author} {\bibfnamefont {B.}~\bibnamefont {Vermersch}},
  \bibinfo {author} {\bibfnamefont {P.}~\bibnamefont {Hauke}}, \bibinfo
  {author} {\bibfnamefont {H.}~\bibnamefont {Pichler}},\ and\ \bibinfo {author}
  {\bibfnamefont {P.}~\bibnamefont {Zoller}},\ }\bibfield  {title} {\bibinfo
  {title} {Non-markovian dynamics in chiral quantum networks with spins and
  photons},\ }\href {https://doi.org/10.1103/PhysRevA.93.062104} {\bibfield
  {journal} {\bibinfo  {journal} {Phys. Rev. A}\ }\textbf {\bibinfo {volume}
  {93}},\ \bibinfo {pages} {062104} (\bibinfo {year} {2016})}\BibitemShut
  {NoStop}%
\bibitem [{\citenamefont {Casulleras}\ \emph {et~al.}(2022)\citenamefont
  {Casulleras}, \citenamefont {Knauer}, \citenamefont {Wang}, \citenamefont
  {Romero-Isart}, \citenamefont {Chumak},\ and\ \citenamefont
  {Gonzalez-Ballestero}}]{casulleras2022generation}%
  \BibitemOpen
  \bibfield  {author} {\bibinfo {author} {\bibfnamefont {S.}~\bibnamefont
  {Casulleras}}, \bibinfo {author} {\bibfnamefont {S.}~\bibnamefont {Knauer}},
  \bibinfo {author} {\bibfnamefont {Q.}~\bibnamefont {Wang}}, \bibinfo {author}
  {\bibfnamefont {O.}~\bibnamefont {Romero-Isart}}, \bibinfo {author}
  {\bibfnamefont {A.~V.}\ \bibnamefont {Chumak}},\ and\ \bibinfo {author}
  {\bibfnamefont {C.}~\bibnamefont {Gonzalez-Ballestero}},\ }\bibfield  {title}
  {\bibinfo {title} {Generation of spin-wave pulses by inverse design},\ }\href
  {https://doi.org/10.48550/arXiv.2209.06608} {\bibfield  {journal} {\bibinfo
  {journal} {arXiv:2209.06608}\ } (\bibinfo {year} {2022})}\BibitemShut
  {NoStop}%
\bibitem [{\citenamefont {Tufarelli}\ \emph {et~al.}(2014)\citenamefont
  {Tufarelli}, \citenamefont {Kim},\ and\ \citenamefont
  {Ciccarello}}]{tufarelli2014non}%
  \BibitemOpen
  \bibfield  {author} {\bibinfo {author} {\bibfnamefont {T.}~\bibnamefont
  {Tufarelli}}, \bibinfo {author} {\bibfnamefont {M.~S.}\ \bibnamefont {Kim}},\
  and\ \bibinfo {author} {\bibfnamefont {F.}~\bibnamefont {Ciccarello}},\
  }\bibfield  {title} {\bibinfo {title} {Non-markovianity of a quantum emitter
  in front of a mirror},\ }\href {https://doi.org/10.1103/PhysRevA.90.012113}
  {\bibfield  {journal} {\bibinfo  {journal} {Phys. Rev. A}\ }\textbf {\bibinfo
  {volume} {90}},\ \bibinfo {pages} {012113} (\bibinfo {year}
  {2014})}\BibitemShut {NoStop}%
\bibitem [{\citenamefont {Laine}\ \emph {et~al.}(2010)\citenamefont {Laine},
  \citenamefont {Piilo},\ and\ \citenamefont {Breuer}}]{laine2010measure}%
  \BibitemOpen
  \bibfield  {author} {\bibinfo {author} {\bibfnamefont {E.-M.}\ \bibnamefont
  {Laine}}, \bibinfo {author} {\bibfnamefont {J.}~\bibnamefont {Piilo}},\ and\
  \bibinfo {author} {\bibfnamefont {H.-P.}\ \bibnamefont {Breuer}},\ }\bibfield
   {title} {\bibinfo {title} {Measure for the non-markovianity of quantum
  processes},\ }\href {https://doi.org/10.1103/PhysRevA.81.062115} {\bibfield
  {journal} {\bibinfo  {journal} {Phys. Rev. A}\ }\textbf {\bibinfo {volume}
  {81}},\ \bibinfo {pages} {062115} (\bibinfo {year} {2010})}\BibitemShut
  {NoStop}%
\bibitem [{\citenamefont {Pellizzari}(1997)}]{Pellizzari1997}%
  \BibitemOpen
  \bibfield  {author} {\bibinfo {author} {\bibfnamefont {T.}~\bibnamefont
  {Pellizzari}},\ }\bibfield  {title} {\bibinfo {title} {{Quantum Networking
  with Optical Fibres}},\ }\href {https://doi.org/10.1103/physrevlett.79.5242}
  {\bibfield  {journal} {\bibinfo  {journal} {Phys. Rev. Lett.}\ }\textbf
  {\bibinfo {volume} {79}},\ \bibinfo {pages} {5242} (\bibinfo {year}
  {1997})}\BibitemShut {NoStop}%
\bibitem [{\citenamefont {Chen}\ \emph {et~al.}(2007)\citenamefont {Chen},
  \citenamefont {Ye}, \citenamefont {Lin}, \citenamefont {Du},\ and\
  \citenamefont {Lin}}]{Chen2007}%
  \BibitemOpen
  \bibfield  {author} {\bibinfo {author} {\bibfnamefont {L.-B.}\ \bibnamefont
  {Chen}}, \bibinfo {author} {\bibfnamefont {M.-Y.}\ \bibnamefont {Ye}},
  \bibinfo {author} {\bibfnamefont {G.-W.}\ \bibnamefont {Lin}}, \bibinfo
  {author} {\bibfnamefont {Q.-H.}\ \bibnamefont {Du}},\ and\ \bibinfo {author}
  {\bibfnamefont {X.-M.}\ \bibnamefont {Lin}},\ }\bibfield  {title} {\bibinfo
  {title} {{Generation of entanglement via adiabatic passage}},\ }\href
  {https://doi.org/10.1103/PhysRevA.76.062304} {\bibfield  {journal} {\bibinfo
  {journal} {Phys. Rev. A}\ }\textbf {\bibinfo {volume} {76}},\ \bibinfo
  {pages} {062304} (\bibinfo {year} {2007})}\BibitemShut {NoStop}%
\bibitem [{\citenamefont {Ye}\ \emph {et~al.}(2008)\citenamefont {Ye},
  \citenamefont {Zhong},\ and\ \citenamefont {Zheng}}]{Ye2008}%
  \BibitemOpen
  \bibfield  {author} {\bibinfo {author} {\bibfnamefont {S.-Y.}\ \bibnamefont
  {Ye}}, \bibinfo {author} {\bibfnamefont {Z.-R.}\ \bibnamefont {Zhong}},\ and\
  \bibinfo {author} {\bibfnamefont {S.-B.}\ \bibnamefont {Zheng}},\ }\bibfield
  {title} {\bibinfo {title} {{Deterministic generation of three-dimensional
  entanglement for two atoms separately trapped in two optical cavities}},\
  }\href {https://doi.org/10.1103/PhysRevA.77.014303} {\bibfield  {journal}
  {\bibinfo  {journal} {Phys. Rev. A}\ }\textbf {\bibinfo {volume} {77}},\
  \bibinfo {pages} {014303} (\bibinfo {year} {2008})}\BibitemShut {NoStop}%
\bibitem [{\citenamefont {Clader}(2014)}]{Clader2014}%
  \BibitemOpen
  \bibfield  {author} {\bibinfo {author} {\bibfnamefont {B.~D.}\ \bibnamefont
  {Clader}},\ }\bibfield  {title} {\bibinfo {title} {Quantum networking of
  microwave photons using optical fibers},\ }\href
  {https://doi.org/10.1103/PhysRevA.90.012324} {\bibfield  {journal} {\bibinfo
  {journal} {Phys. Rev. A}\ }\textbf {\bibinfo {volume} {90}},\ \bibinfo
  {pages} {012324} (\bibinfo {year} {2014})}\BibitemShut {NoStop}%
\bibitem [{\citenamefont {Tian}(2012)}]{Tian2012}%
  \BibitemOpen
  \bibfield  {author} {\bibinfo {author} {\bibfnamefont {L.}~\bibnamefont
  {Tian}},\ }\bibfield  {title} {\bibinfo {title} {{Adiabatic State Conversion
  and Pulse Transmission in Optomechanical Systems}},\ }\href
  {https://doi.org/10.1103/PhysRevLett.108.153604} {\bibfield  {journal}
  {\bibinfo  {journal} {Phys. Rev. Lett.}\ }\textbf {\bibinfo {volume} {108}},\
  \bibinfo {pages} {153604} (\bibinfo {year} {2012})}\BibitemShut {NoStop}%
\bibitem [{\citenamefont {Pechal}\ \emph {et~al.}(2014)\citenamefont {Pechal},
  \citenamefont {Huthmacher}, \citenamefont {Eichler}, \citenamefont
  {Zeytino\ifmmode~\breve{g}\else \u{g}\fi{}lu}, \citenamefont {Abdumalikov},
  \citenamefont {Berger}, \citenamefont {Wallraff},\ and\ \citenamefont
  {Filipp}}]{Pechal2013}%
  \BibitemOpen
  \bibfield  {author} {\bibinfo {author} {\bibfnamefont {M.}~\bibnamefont
  {Pechal}}, \bibinfo {author} {\bibfnamefont {L.}~\bibnamefont {Huthmacher}},
  \bibinfo {author} {\bibfnamefont {C.}~\bibnamefont {Eichler}}, \bibinfo
  {author} {\bibfnamefont {S.}~\bibnamefont {Zeytino\ifmmode~\breve{g}\else
  \u{g}\fi{}lu}}, \bibinfo {author} {\bibfnamefont {A.~A.}\ \bibnamefont
  {Abdumalikov}}, \bibinfo {author} {\bibfnamefont {S.}~\bibnamefont {Berger}},
  \bibinfo {author} {\bibfnamefont {A.}~\bibnamefont {Wallraff}},\ and\
  \bibinfo {author} {\bibfnamefont {S.}~\bibnamefont {Filipp}},\ }\bibfield
  {title} {\bibinfo {title} {Microwave-controlled generation of shaped single
  photons in circuit quantum electrodynamics},\ }\href
  {https://doi.org/10.1103/PhysRevX.4.041010} {\bibfield  {journal} {\bibinfo
  {journal} {Phys. Rev. X}\ }\textbf {\bibinfo {volume} {4}},\ \bibinfo {pages}
  {041010} (\bibinfo {year} {2014})}\BibitemShut {NoStop}%
\bibitem [{\citenamefont {Zeytino\ifmmode~\breve{g}\else \u{g}\fi{}lu}\ \emph
  {et~al.}(2015)\citenamefont {Zeytino\ifmmode~\breve{g}\else \u{g}\fi{}lu},
  \citenamefont {Pechal}, \citenamefont {Berger}, \citenamefont {Abdumalikov},
  \citenamefont {Wallraff},\ and\ \citenamefont {Filipp}}]{Zeytinoglu2015}%
  \BibitemOpen
  \bibfield  {author} {\bibinfo {author} {\bibfnamefont {S.}~\bibnamefont
  {Zeytino\ifmmode~\breve{g}\else \u{g}\fi{}lu}}, \bibinfo {author}
  {\bibfnamefont {M.}~\bibnamefont {Pechal}}, \bibinfo {author} {\bibfnamefont
  {S.}~\bibnamefont {Berger}}, \bibinfo {author} {\bibfnamefont {A.~A.}\
  \bibnamefont {Abdumalikov}}, \bibinfo {author} {\bibfnamefont
  {A.}~\bibnamefont {Wallraff}},\ and\ \bibinfo {author} {\bibfnamefont
  {S.}~\bibnamefont {Filipp}},\ }\bibfield  {title} {\bibinfo {title}
  {Microwave-induced amplitude- and phase-tunable qubit-resonator coupling in
  circuit quantum electrodynamics},\ }\href
  {https://doi.org/10.1103/PhysRevA.91.043846} {\bibfield  {journal} {\bibinfo
  {journal} {Phys. Rev. A}\ }\textbf {\bibinfo {volume} {91}},\ \bibinfo
  {pages} {043846} (\bibinfo {year} {2015})}\BibitemShut {NoStop}%
\bibitem [{\citenamefont {Cirac}\ and\ \citenamefont
  {Zoller}(1995)}]{Cirac1995}%
  \BibitemOpen
  \bibfield  {author} {\bibinfo {author} {\bibfnamefont {J.~I.}\ \bibnamefont
  {Cirac}}\ and\ \bibinfo {author} {\bibfnamefont {P.}~\bibnamefont {Zoller}},\
  }\href@noop {} {\emph {\bibinfo {title} {{Quantum Computations with Cold
  Trapped Ions}}}},\ \bibinfo {type} {Tech. Rep.}\ (\bibinfo {year}
  {1995})\BibitemShut {NoStop}%
\bibitem [{\citenamefont {Gardiner}\ and\ \citenamefont
  {Collett}(1985)}]{Gardiner1985}%
  \BibitemOpen
  \bibfield  {author} {\bibinfo {author} {\bibfnamefont {C.~W.}\ \bibnamefont
  {Gardiner}}\ and\ \bibinfo {author} {\bibfnamefont {M.~J.}\ \bibnamefont
  {Collett}},\ }\bibfield  {title} {\bibinfo {title} {{Input and output in
  damped quantum systems: Quantum stochastic differential equations and the
  master equation}},\ }\href {https://doi.org/10.1103/PhysRevA.31.3761}
  {\bibfield  {journal} {\bibinfo  {journal} {Phys. Rev. A}\ }\textbf {\bibinfo
  {volume} {31}},\ \bibinfo {pages} {3761} (\bibinfo {year}
  {1985})}\BibitemShut {NoStop}%
\bibitem [{\citenamefont {Morin}\ \emph {et~al.}(2019)\citenamefont {Morin},
  \citenamefont {K\"orber}, \citenamefont {Langenfeld},\ and\ \citenamefont
  {Rempe}}]{morin2019deterministic}%
  \BibitemOpen
  \bibfield  {author} {\bibinfo {author} {\bibfnamefont {O.}~\bibnamefont
  {Morin}}, \bibinfo {author} {\bibfnamefont {M.}~\bibnamefont {K\"orber}},
  \bibinfo {author} {\bibfnamefont {S.}~\bibnamefont {Langenfeld}},\ and\
  \bibinfo {author} {\bibfnamefont {G.}~\bibnamefont {Rempe}},\ }\bibfield
  {title} {\bibinfo {title} {Deterministic shaping and reshaping of
  single-photon temporal wave functions},\ }\href
  {https://doi.org/10.1103/PhysRevLett.123.133602} {\bibfield  {journal}
  {\bibinfo  {journal} {Phys. Rev. Lett.}\ }\textbf {\bibinfo {volume} {123}},\
  \bibinfo {pages} {133602} (\bibinfo {year} {2019})}\BibitemShut {NoStop}%
\bibitem [{\citenamefont {Gorshkov}\ \emph {et~al.}(2007)\citenamefont
  {Gorshkov}, \citenamefont {Andr\'e}, \citenamefont {Lukin},\ and\
  \citenamefont {S\o{}rensen}}]{gorshkov2007photon}%
  \BibitemOpen
  \bibfield  {author} {\bibinfo {author} {\bibfnamefont {A.~V.}\ \bibnamefont
  {Gorshkov}}, \bibinfo {author} {\bibfnamefont {A.}~\bibnamefont {Andr\'e}},
  \bibinfo {author} {\bibfnamefont {M.~D.}\ \bibnamefont {Lukin}},\ and\
  \bibinfo {author} {\bibfnamefont {A.~S.}\ \bibnamefont {S\o{}rensen}},\
  }\bibfield  {title} {\bibinfo {title} {Photon storage in
  $\ensuremath{\Lambda}$-type optically dense atomic media. i. cavity model},\
  }\href {https://doi.org/10.1103/PhysRevA.76.033804} {\bibfield  {journal}
  {\bibinfo  {journal} {Phys. Rev. A}\ }\textbf {\bibinfo {volume} {76}},\
  \bibinfo {pages} {033804} (\bibinfo {year} {2007})}\BibitemShut {NoStop}%
\bibitem [{\citenamefont {Gardiner}\ and\ \citenamefont
  {Zoller}(2015)}]{GardinerUltracoldII}%
  \BibitemOpen
  \bibfield  {author} {\bibinfo {author} {\bibfnamefont {C.}~\bibnamefont
  {Gardiner}}\ and\ \bibinfo {author} {\bibfnamefont {P.}~\bibnamefont
  {Zoller}},\ }\href {https://doi.org/10.1142/9781783266784} {\emph {\bibinfo
  {title} {{The Quantum World of Ultra-Cold Atoms and Light Book II: The
  Physics of Quantum-Optical Devices}}}}\ (\bibinfo  {publisher} {World
  Scientific},\ \bibinfo {year} {2015})\BibitemShut {NoStop}%
\bibitem [{\citenamefont {Ripoll}(2022)}]{ripoll2022quantum}%
  \BibitemOpen
  \bibfield  {author} {\bibinfo {author} {\bibfnamefont {J.~J.~G.}\
  \bibnamefont {Ripoll}},\ }\href@noop {} {\emph {\bibinfo {title} {Quantum
  Information and Quantum Optics with Superconducting Circuits}}}\ (\bibinfo
  {publisher} {Cambridge University Press},\ \bibinfo {year}
  {2022})\BibitemShut {NoStop}%
\bibitem [{\citenamefont {{David M. Pozar}}(2012)}]{Pozar}%
  \BibitemOpen
  \bibfield  {author} {\bibinfo {author} {\bibnamefont {{David M. Pozar}}},\
  }\href@noop {} {\emph {\bibinfo {title} {{Microwave Engineering}}}}\
  (\bibinfo  {publisher} {John Wiley \& Sons},\ \bibinfo {year}
  {2012})\BibitemShut {NoStop}%
\end{thebibliography}%

\end{document}